\documentclass{aa} 
\usepackage{graphicx} 
\usepackage{txfonts} 
\usepackage{natbib} 
\bibpunct{(}{)}{;}{a}{}{,} 

\begin{document} 
\title{\ion{Ca}{ii} and \ion{Na}{i} absorption signatures from extraplanar gas in the halo of the Milky Way} 
\titlerunning{\ion{Ca}{ii} and \ion{Na}{i} absorption signatures from gas in the MW halo} 
\subtitle{} 
\author{N. Ben Bekhti\inst{1}      
\and P. Richter\inst{2}    
\and T. Westmeier\inst{3}
\and M. T. Murphy\inst{4}
} 
\authorrunning{Ben Bekhti et al.}
\offprints{N. Ben Bekhti} 
\institute{Argelander-Institut f\"ur Astronomie, 
Universit\"{a}t Bonn, Auf dem H\"{u}gel 71, 53121 Bonn, Germany\\
\email{nbekhti@astro.uni-bonn.de}   
\and Institut f\"{u}r Physik und Astronomie, 
Universit\"{a}t Potsdam, Haus 28, Karl-Liebknecht-Str.\,24/25,
14476 Potsdam, Germany\\ 
\email{prichter@astro.physik.uni-potsdam.de} 
\and Australia Telescope
National Facility, PO Box 76, Epping NSW 1710, Australia\\ \email{tobias.westmeier@csiro.au}
\and  Centre for Astrophysics \& Supercomputing, Swinburne
University of Technology, Hawthorn, Victoria 3122, Australia \\ \email{mmurphy@swin.edu.au}}

\date{Received xxxx; accepted month xxx} 

\abstract{}{We analyse absorption characteristics and physical conditions of extraplanar intermediate- and high-velocity gas to study the distribution of the neutral and weakly ionised Milky Way halo gas and its relevance for the evolution of the Milky Way and other spiral galaxies.}
{We combine optical absorption line measurements of \ion{Ca}{ii}/\ion{Na}{i} and
21\,cm emission line observations of \ion{H}{i} along 103 extragalactic lines of sight towards quasars (QSOs) and active galactic nuclei (AGN). The archival optical spectra were obtained with the Ultraviolet and Visual Echelle Spectrograph (UVES) at the ESO Very Large Telescope, while the 21\,cm \ion{H}{i} observations were carried  out using the 100-m radio telescope at Effelsberg.}
{The analysis of the UVES spectra shows that single and multi-component \ion{Ca}{ii}/\ion{Na}{i} absorbers at intermediate and high velocities are present in about 35 percent of the sight lines, indicating the presence of neutral extraplanar gas structures. In some cases the \ion{Ca}{ii}/\ion{Na}{i} absorption is connected with \ion{H}{i} 21\,cm intermediate- or high-velocity gas with \ion{H}{i} column densities in the range of $10^{18}$ to $10^{20}\,\mathrm{cm}^{-2}$ (i.e., the classical IVCs and HVCs), while other \ion{Ca}{ii}/\ion{Na}{i} absorbers show no associated \ion{H}{i} emission. The observed \ion{H}{i} line widths vary from $\Delta v_\mathrm{FWHM}=3.2$\,km\,s$^{-1}$ to $32.0$\,km\,s$^{-1}$ indicating a range of upper gas temperature limits of 250\,K up to about 22500\,K.}
{Our study suggests that the Milky Way halo is filled with a large number of neutral gaseous structures whose high column density tail represents the population of common \ion{H}{i} high-velocity clouds seen in 21\,cm surveys. The \ion{Ca}{ii} column density distribution follows a power-law $f(N)=CN^{\beta}$ with a slope of $\beta \approx -1.6$, thus comparable to the distribution found for intervening metal-line systems toward QSOs. Many of the statistical and physical properties of the \ion{Ca}{ii} absorbers resemble those of strong ($W_\mathrm{\lambda 2796}>0.3\,\AA{}$) \ion{Mg}{ii} absorbing systems observed in the circumgalactic environment of other galaxies, suggesting that both absorber populations may be closely related.} 

\keywords{Galaxy: halo -- ISM: structure -- quasars: absorption lines -- Galaxies: halo} 
\maketitle 

\section{Introduction}\label{Introduction}


Spiral galaxies are surrounded by large gaseous halos. That the disk of the Milky Way has a hot envelope was first proposed by \citet{spitzer56}. Spitzer considered such a ``Galactic Corona'' to explain spectroscopic observations that have been made earlier by \citet{adams49} and \citet{muench52}. In recent years, great instrumental progress has been made to measure extraplanar gas structures around other galaxies, as well. It became clear, that the gaseous halos represent the interface between the condensed galactic discs and the surrounding intergalactic medium (IGM) \citep[e.g.,][ and references therein]{savage_massa87, majewski04, fraternalietal_07}.  The properties of this medium around galaxies presumably are determined by both, the accretion of metal-poor gaseous matter from intergalactic space onto the galactic disc, as well as the outflow of metal-enriched gas caused by star formation activity within the galaxy \citep[e.g.,][ and references therein]{sembach_wakker_savage_richter_etal03, fraternaliandbinney06}.

%

Gas in the halos and intergalactic environment of galaxies leaves its imprint in the spectra of distant quasars (QSOs) in the form of hydrogen- and metal-line absorption \citep[for a recent review, see][]{richterp06}. Therefore, QSO absorption spectroscopy has become a powerful method to study the physical properties, the kinematics, and the spatial distribution of gas in the halos of galaxies over a large range of column densities at low and high redshifts. The analysis of intervening \ion{Mg}{ii} and \ion{C}{iv} absorption line systems \citep[e.g.,][]{charltonetal00, dingetal03, masieroetal05, boucheetal06} and their relation to galactic structures suggest rather complex absorption characteristics of these ions indicating the multi-phase nature of gas in the outskirts of galaxies with density and temperature ranges spanning several order of magnitudes. Stronger intervening low-ion absorbers (e.g., strong \ion{Mg}{ii} systems) preferentially arise at low impact parameters ($< 35\,h^{-1}$\,kpc) of intervening galaxies, while weaker \ion{Mg}{ii} systems and high-ion absorbers (e.g., \ion{C}{iv} systems) apparently are often located at larger distances up to $\sim 100\,h^{-1}$\,kpc  (Churchill et al.\,1999; Milutinovic et al.\,2006). Due to the lack of additional information the exact nature and origin of the various circumgalactic absorber populations is not yet fully understood. Most likely, gas outflow and infall processes both contribute to the complex absorption pattern observed.

Our own Galaxy also is surrounded by large amounts of neutral and ionised gas \citep{richterp06}. Most prominent are the so-called intermediate- and high-velocity clouds \citep[IVCs, HVCs,][] {mulleroortraimond63} which represent clouds of neutral atomic hydrogen seen in 21\,cm emission at radial velocities inconsistent with a simple model of Galactic disk rotation. Most important for our understanding of the nature of IVCs and HVCs and their role for the evolution of the Milky Way is the determination of accurate metal abundances and distances of these clouds.
Metallicity measurements are particularly important to learn about the origin 
of IVCs/HVCs. The metal abundances of some IVCs/HVCs have been determined by absorption line measurements along several lines of sight \citep[see][]{wakker01, richterp06}. The results show that the metallicities are varying between $\sim 0.1$ and $\sim 1.0$ solar. This wide range of metallicities suggests that many IVCs and HVCs cannot have a common origin. In fact, it is know widely accepted that various different processes contribute to the neutral gas flow in the Milky Way halo including the Galactic fountain \citep{shapirofield76, bregman80, shapiro_benjamin91}, the accretion of gas from surrounding satellite galaxies \citep[e.g., Magellanic Stream, ][]{mathewson74}, and the infall of metal-poor gas from the intergalactic medium (e.g., Wakker et al.\,1999). 

To determine the total mass of the gas that is falling toward the Milky Way disk in form of IVCs and HVCs a reliable distance estimate of these clouds is required. Measuring the distances of IVCs/HVCs is very difficult, however. The most reliable method to infer a distance bracket requires high-resolution spectra of stars with known distances in which IVCs/HVCs appear in absorption. The problem is the limited number of suitable background stars. The distance estimates of IVCs and HVCs around the Milky Way \citep[e.g.,][]{Sembachetal91, vanWoerden99, wakker01, thom2006, wakker_york_howketal07, wakkeryorkwilhelmetal08} indicate that most IVCs are relatively nearby objects with distances of $d< 2$\,kpc while the majority of the HVCs are more distant clouds, located in the halo of the Milky Way with distances of $5<d<50$\,kpc.
These numbers imply that the typical mass-accretion rate from IVCs and HVC are on the order of one solar mass per year.

While there are a large number of recent absorption studies on the nature of IVCs and HVCs and their role for the evolution of the Milky Way, relatively little effort has been made to investigate the connection between the Galactic population of IVCs and HVCs and the distribution and nature of intervening metal-absorption systems from galaxy halos seen in QSO spectra.
In fact, almost all recent absorption studies of IVCs and HVCs were carried out in the FUV to study in detail metal abundances and ionisation conditions of  halo clouds using the many available transitions of low and high ions in the ultraviolet regime (e.g., Richter et al.\,2001). These studies were designed as follow-up absorption observations of known IVCs and HVCs, thus providing an 21\,cm {\it emission-selected} data set. 
However, to statistically compare the absorption characteristics of the extraplanar Galactic halo structures with the properties of intervening metal-absorption systems towards QSOs one requires an {\it absorption-selected} data set of IVCs and HVCs. Since in the UV band there are currently only a very limited number ($<50$) of high-quality spectra available such a statistical comparison can be done best in the optical regime where a large number of high-quality spectra of low- and high-redshift QSOs are available.

In this paper we discuss low-column density extraplanar structures which are most likely located in the environment of the Milky Way detected in optical \ion{Ca}{ii} and \ion{Na}{i} absorption towards quasars along 103 sight lines through the Milky Way halo. Our study allows us to directly compare the observed absorption column-density distribution of gas in the Milky Way halo with the overall column-density distribution of intervening metal absorbers toward QSOs and AGN. Moreover, our study enables us to identify neutral and ionised absorption structures at low gas column densities and small angular extent that remain unseen in the large 21\,cm IVC and HVC all-sky surveys but that possibly have a considerable absorption cross section \citep[see][]{richterwestmeierbruens05}. We supplement our absorption-line data with new \ion{H}{i} 21\,cm observations to investigate the relation between 
intermediate- and high-velocity \ion{Ca}{ii} absorption features and halo 21\,cm emission. The study presented here discusses the first results of our analysis of the physical and statistical properties of the detected absorption and emission features. Additional absorption-line observations as well as follow-up \ion{H}{i} synthesis observations of some of the absorbers are under way and will be presented in a subsequent paper (Ben Bekhti et al., 2008; in preparation). 

Our paper is organised as follows. In Section\,\ref{data} we describe the data acquisition and data reduction. In Section\,\ref{results} the results of the analysis of the UVES and the Effelsberg data of 103 sight lines in the direction of several quasars are presented. A statistical investigation (column density distribution functions, distribution of $b$-values, deviation velocities, etc.)  of the data is presented in Section\,\ref{stat.properties}. In Section\,\ref{disscussion} we discuss the results regarding multiple absorption lines and corresponding emission line measurements, the determination of metallicities and a probable association of the gas with known IVC and HVC complexes. In Section\,\ref{summary} we summarise our results and discuss future observations that will be required to determine the metallicities of the gas, important to understand the nature and origin of these intermediate- and high-velocity absorption features. 

\section{Data acquisition and reduction}\label{data} 
\subsection{UVES data} 

The data basis of our analysis are 103 optical spectra of high-redshift QSOs and AGN obtained with the Ultraviolet and Visual Echelle Spectrograph (UVES) between 1999 and 2004 at the ESO Very Large Telescope. These spectra, observed for various
purposes by various groups, are publically available in the ESO data archive.\footnote{http://archive.eso.org/wdb/wdb/eso/uves/form} The spectra used in this study have a spectral resolution of $R \approx 40000-60000$, corresponding to approximately $6.6\,\mathrm{km\,s}^{-1}$\,FWHM. For the normalised spectra the signal-to-noise ratio per resolution element 

\begin{equation}
S/N_\mathrm{r}=\sqrt{\frac{\Delta \lambda_\mathrm{r}}{\Delta \lambda_\mathrm{p}}}\frac{1}{\sigma_\mathrm{p}^\mathrm{rms}}
\end{equation} 

at the two \ion{Ca}{ii} lines near 4000\,\AA{} is between $10$ and $190$ with a typical value (median) of about $90$. The quantity $\Delta \lambda_\mathrm{r,p}$ are the wavelength separations per resolution and pixel element, respectively. The noise per pixel is given by $\sigma_\mathrm{p}^\mathrm{rms}$.  A detailed description of the UVES instrument is given by \citet{dekkeretal.00}.

Part of the raw data were reduced with the UVES pipeline implemented in the
\textsc{ESO-Midas} software-package as part of the data reduction process of the UVES Large Programme \citep{bergeronpetitjeanetal04}. The pipeline reduction includes flat-fielding, bias- and sky-subtraction, and a relative wavelength calibration. The data were then normalised by a continuum fit using high-order polynomials. The remaining data were reduced and normalised as part of the UVES
Spectral Quasar Absorption Database (SQUAD). A version of the UVES pipeline, modified to improve the flux extraction and wavelength calibration, was used to extract the echelle orders. These were combined using the custom-written code, {\sc uvespopler},\footnote{Available at http://www.ast.cam.ac.uk/$\sim$mim/UVES\_popler.html} with inverse-variance weighting and a cosmic ray rejection algorithm, to form a single spectrum with a dispersion of $2.5\,\mathrm{km\,s}^{-1}\,\mathrm{pixel}^{-1}$.

As the two \ion{Ca}{ii} lines at 3934.77\,\AA\ and 3969.59\,\AA\ represent a doublet, we define a detection limit for the high-velocity \ion{Ca}{ii} absorption of $4 \sigma$ for the stronger line at $\lambda = 3934.77\,\AA{}$ and a $2 \sigma$ limit for the weaker line at $\lambda = 3969.59\,\AA{}$. For the median (minimal, maximal) $S/N_\mathrm{r}$ of about 90 (10, 190) the former limit corresponds to an equivalent width limit of $W_{\lambda}
\approx 4\,\mathrm{m}\AA{}$ ($50\,\mathrm{m}\AA{}$, $2\,\mathrm{m}\AA{}$). These limits can be converted to column density detection limits on the linear part of the curve of growth using

\begin{equation}
 N=\frac{W_\lambda\, \mathrm{m}_\mathrm{e}c^2}{\pi\, \mathrm{e}^2\,\lambda^2\,f}.
\end{equation} 

The spectral features were analysed via Voigt-profile fitting using the \textsc{Fitlyman} package in \textsc{Midas}, which, among other parameters, delivers column densities and Doppler parameters ($b$-values).

\begin{figure*} 
\includegraphics[width=\textwidth,clip]{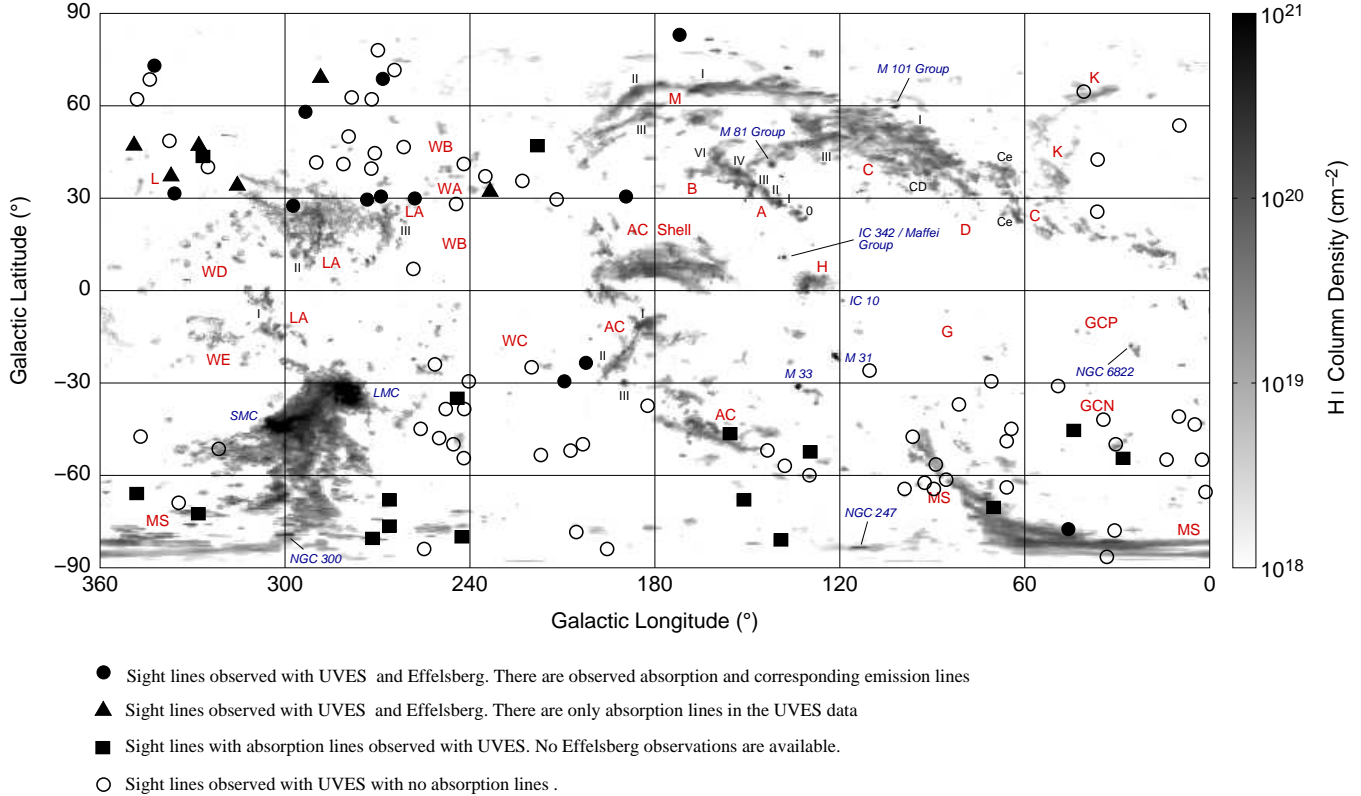} 
\caption{HVC-all-sky map created by \citet{westmeier07} derived from the data of
the LAB Survey (Kalberla et al. 2005). The different symbols mark the positions
of 103 sight lines that were observed with UVES. Along 35 lines of sight we
detect \ion{Ca}{ii}/\ion{Na}{ii} absorption components. For 19 of them we have obtained additional \ion{H}{i} observations with the Effelsberg 100-m telescope.} 
\label{fig_hvc_allskymap} 
\end{figure*}

\subsection{Effelsberg data} 

The follow-up \ion{H}{i} 21\,cm observations were obtained in 2006 with the 100-m radio telescope at Effelsberg for which at 21\,cm wavelength the half-power beam width (HPBW) is $9\arcmin$. For our observations, the velocity resolution is about $0.5\,\mathrm{kms}^{-1}$, and the rms is a few times $10^{-2}$\,K, while the integration time per sight line was about 25\,min. The spectra were analysed using the \textsc{Gildas} tool \textsc{Class}. In all spectra polynomial baselines of up to $4^\mathrm{th}$ order were fitted. For this purpose, windows were set individually around the line emission. All data within these windows were not considered for the fit. After the baseline fit Gaussian functions were fitted to the spectral lines.

\subsection{The Leiden/Argentine/Bonn Survey (LAB)}

In addition to our Effelsberg observations we used archival data from the Leiden-Argentine-Bonn (LAB) all-sky \ion{H}{i} 21\,cm survey \citep{kalberlaetal05} for those 
sight lines where we do not have Effelsberg spectra to search for \ion{H}{i} emission related to the optical absorption.
 
The LAB survey is a combination of the Leiden/Dwingeloo Survey \citep[LDS,][]{hartmannburton97}, covering the sky north of $\delta=-30^\circ$, and the Instituto Argentino de Radio\-astronom\'{i}a Survey \citep[IAR,][]{arnalbajajaetal00, bajajaarnaletal05} which covers the sky south of
$\delta=-25^\circ$. The HPBW of the LAB survey is about $36 \arcmin$. The LSR 
velocity coverage is in the range of $-450\,\mathrm{kms}^{-1}$ to $+400
\,\mathrm{kms}^{-1}$, at a velocity resolution of $1.3\,\mathrm{kms}^{-1}$. The entire LAB  data was corrected for stray radiation at the Argelander-Institut f\"{u}r Astronomie in Bonn.    

\section{Observational results and sample selection}\label{results} 

Fig.\,\ref{fig_hvc_allskymap} shows an all-sky HVC map which was created by \citet{westmeier07} based on the data of the LAB survey \citep{kalberlaetal05}. On the one hand, several extended \ion{H}{i} HVC complexes cover the sky \citep{wakker91}, some of which are spanning tens of degrees. On the other hand, numerous isolated and compact HVCs can be seen all over the sky \citep{braunburton99, Putmanetal02, deheijbraunburton02}. However, most of these structures are not resolved with the $36\arcmin$ beam of the LAB survey. The symbols in Fig.\,\ref{fig_hvc_allskymap} mark the positions of the 103~sight lines that were observed with UVES.  

For the further analysis we have transformed all spectra to the Local Standard of Rest (LSR) velocity scale. Note most of the gas near zero velocities is located in the Galactic disk. To separate low-velocity gas in the disk from extraplanar intermediate- and high-velocity gas in the halo we have used the concept of the so-called deviation velocity together with a kinematic model for the Milky Way developed by \citet{kalberla03}. With this model we can determine whether an observed velocity in a given direction is expected for interstellar gas participating in the Galactic disk rotation or not. According to \citet{wakker91} the deviation velocity is defined by

\begin{equation}
v_\mathrm{dev}=\left\{
\begin{array}{ccl}
v_\mathrm{LSR} - v_\mathrm{min} & ~ & \mathrm{if}~v_\mathrm{LSR} < 0\\
v_\mathrm{LSR} - v_\mathrm{max} & ~ & \mathrm{if}~v_\mathrm{LSR} > 0
\end{array}
\right\},
\end{equation}
where $v_\mathrm{min}$ and $v_\mathrm{max}$ are given by the rotation model of
\citet{kalberla03}.
We consider all absorption lines with $v_\mathrm{dev} > 0$ (if $v_\mathrm{LSR} > 0$) or $v_\mathrm{dev} < 0$ (if $v_\mathrm{LSR} < 0$) as intermediate- or high-velocity gas. To distinguish between IVCs and HVCs we use the  definition of \citet{wakker91} in which a cloud is defined as an IVC, if $ \vert v_\mathrm{dev}\vert \leq 50$\, \textrm{km\,s}$^{-1}$ and as an HVC if $ \vert v_\mathrm{dev}\vert > 50$\,\textrm{km\,s}$^{-1}$. Using these selection criteria for the 103 UVES spectra we detect 55 \ion{Ca}{ii} and 20 \ion{Na}{i} halo absorption components along 35~lines of sight at intermediate and high velocities. These 35 sighlines are indicated in Fig.\,\ref{fig_hvc_allskymap} with the filled symbols. For 19 of these sight lines we have carried out deep \ion{H}{i} 21\,cm observations with the Effelsberg 100-m telescope to search for neutral hydrogen structures that are associated with the absorption systems.

A complete presentation of the 35 obtained spectra with optical absorption of \ion{Ca}{ii} $\lambda$3934.77\,$\AA{}$, \ion{Ca}{ii} $\lambda$3969.59\,$\AA{}$, \ion{Na}{i} $\lambda$5891.58\,$\AA{}$, \ion{Na}{i} $\lambda$5897.56\,$\AA{}$ together with the corresponding \ion{H}{i} $\lambda$21\,cm emission profiles observed with the 100-m telescope at Effelsberg is given in the Appendix.

Some of the halo \ion{Ca}{ii} and \ion{Na}{i} absorbers show multiple intermediate- and high-velocity components, indicating the presence of gaseous sub-structures. Furthermore, along 13 sight lines the intermediate- and high-velocity \ion{Ca}{ii} and \ion{Na}{i} absorption is connected with \ion{H}{i} gas with typical column densities in the range of a few times $10^{18}\,\mathrm{cm}^{-2}$ up to $10^{20}\,\mathrm{cm}^{-2}$. The measured \ion{H}{i} line widths vary from $\Delta v_\mathrm{FWHM}=3.2$\,km\,s$^{-1}$ to $32.0$\,km\,s$^{-1}$ giving a range of upper gas temperature limits of 250\,K up to about 22500\,K. Measured column densities (UVES and Effelsberg) and $b$-values for the 35 sight lines are summarised in Tables\,\ref{tab_27_quasars} and~\ref{tab_20_quasars}. 

It is important to note at this point that in many cases there is no high- or intermediate-velocity \ion{H}{i} 21\,cm emission seen in the low-resolution LAB survey, whereas the higher resolution Effelsberg telescope detects emission lines at the corresponding position. This demonstrates how important high-resolution measurements are in view of beam smearing effects and the fact that the \ion{H}{i} column densities of most of the \ion{Ca}{ii} and \ion{Na}{i} absorption components are close to or below the detection limit of large \ion{H}{i} surveys. One sight line (PKS~1448$-$232) with particularly prominent high-velocity \ion{Ca}{ii} and \ion{Na}{i} absorption lines was observed by \citet{richterwestmeierbruens05} with UVES, Effelsberg, and the VLA. The high-resolution VLA \ion{H}{i} data resolves the HVC into several compact, cold clumps. In our sample, 19~sight lines show intermediate- or high-velocity
\ion{Ca}{ii}/\ion{Na}{i} absorption without any corresponding counterparts in the \ion{H}{i} data, suggesting that either the \ion{H}{i} column densities are below the detection limit of the Effelsberg radio telescope or that the diameters of these clouds are very small so that beam-smearing effects make them undetectable. 

\begin{table*}
\caption{Summary of the UVES and Effelsberg measurements for intermediate and high velocity \ion{Ca}{ii} and \ion{Na}{i} absorbers toward 20
quasars. The columns give the name of the quasars, the coordinates, the
velocities in the LSR frame, the logarithm of the \ion{Ca}{ii} column densities,
the $b$-value for the \ion{Ca}{ii} lines, the logarithm of the \ion{Na}{i}
column densities, the $b$-value for the \ion{Na}{i} lines, the \ion{H}{i} column
densities, the $b$-value for the \ion{H}{i} lines and the possibly associated HVC/IVC complex. The $4\sigma$ and
$2\sigma$ detection limit (Section\,\ref{data}) for \ion{Ca}{ii}~$\lambda
3934.77$ and \ion{Ca}{ii}~$\lambda 3969.59$, respectively, corresponds to a
$\log (N_\mathrm{CaII}/\mathrm{cm}^{-2})$ detection limit of $\approx 11$. n/a
means that no Effelsberg data are available for these sight lines.}
\label{tab_27_quasars}
\small
\centering
\begin{tabular}{ccccccccccc}\hline\hline
\rule{0pt}{3ex}Quasar&$l$&$b$&$v_\mathrm{LSR}$&$\log N_\mathrm{CaII}$&$b_\mathrm{CaII}$&$\log N_\mathrm{NaI}$&$b_\mathrm{NaI}$& $\log N_\mathrm{HI}$&$b_\mathrm{HI}$&HVC/IVC\\

\rule{0pt}{3ex}&[$\degr$]&[$\degr$]&[km\,s$^{-1}$]&[$N$ in cm$^{-2}$]&[km\,s$^{-1}$]& [$N$ in cm$^{-2}$]&[km\,s$^{-1}$]&[$N$ in cm$^{-2}$]&[km\,s$^{-1}$]&complex\\[1ex]\hline
\rule{0pt}{3ex} QSO J1232-0224&293.2&60.1&$-$21&11.5&2&11.0&1&19.9&10.2&-\\\hline
QSO B1101-26 &275.0&30.2&199&11.7&6&-&-&-&-&LA\\
&&& $-$17&11.7&6&11.2&3&20.1&6.8&-\\
&&&$-$27&11.3& 4&-&-&-&-&-\\\hline
QSO J0003-2323&49.4&$-$78.6&$-$98& 11.9& 6 &-&-&-&-&MS\\
&&&$-$112& 11.8& 6 &-&-&19.6&19.3&MS\\
&&&$-$126& 11.9& 6 &-&-&-&-&MS\\\hline
QSO B0450-1310B &211.8&$-$32.1& $-$5&-&-&11.5&5&19.8&2.3&- \\
&&&$-$20&-&-&12.0&2&19.3&6.8&- \\\hline
QSO B0515-4414&249.6&$-$35.0&$-$5&11.7&3&9.4&2&n/a&-&-\\
&&&$-$17&10.9&1&-&-&n/a&-&-\\
&&&$-$41&11.3&4&10.8&3&n/a&-&-\\
&&&$-$58&11.3&1&-&-&n/a&-&-\\\hline
QSO B1036-2257&267.4&30.4&$-$24&11.9&7&12.0&1&19.6&2.0&-\\\hline
QSO J1344-1035&323.5&50.2&$-$65& 11.8&4&-&-&-&-&-\\\hline
QSO B0109-353&275.5&$-$81.0&79&12.4&7&11.0&6&n/a&-&-\\  
&&&$-$108&12.4&7&-&-&n/a&-&MS\\
&&&$-$162&12.3&12&-&-&n/a&-&MS\\\hline
QSO B1448-232&335.4&31.7&$-$100&10.9&2&-&-&-&-&L\\ 
&&&$-$130&11.6&9&-&-&-&-&L\\
&&&$-$150&11.6&1&11.8&2&18.9&5.2&L\\
&&&$-$157&11.5&5&-&-&-&-&L\\\hline
QSO B0002-422&332.7&$-$72.4&89&11.8&6&-&-&n/a&-&MS\\\hline
QSO B0122-379&271.9&$-$77.3&40&11.7&4&-&-&n/a&-&MS\\\hline
QSO B1347-2457&319.5&35.8&$-$95&11.5&3&-&-&-&-&-\\\hline
J092913-021446&235.7&33.2&176&11.7&3&-&-&-&-&WA\\\hline
J081331+254503&196.9&28.6&$-$23&-&-&11.1&5&19.8&37.0&-\\\hline
J222756-224302&32.6&$-$57.3&$-$118&12.1&7&-&-&n/a&-&-\\\hline
QSO J0830+2411&200.0&31.9&$-$21&-&-&11.3&3&19.2&7.6&-\\\hline
QSO B0952+179&216.5&48.4&$-$30&-&-&11.6&9&n/a&-&IV Spur\\\hline
QSO J2155-0922&47.5&$-$44.8&$-$166&11.1&1&-&-&n/a&-&GCN\\
&&&$-$209&11.8&6&-&-&n/a&-&GCN\\\hline
QSO J1211+1030&271.7&70.9&76&11.5&2&-&-&-&-&-\\
&&&$-$26&11.6&4&-&-&19.7&13.5&IV Spur?\\\hline
J143649.8-161341&336.6&39.7&76&11.4&3&-&-&-&-&-\\\hline


\end{tabular}
\end{table*}

\begin{table*}
\caption{Summary of the UVES and Effelsberg measurements for intermediate and high velocity \ion{Ca}{ii} and \ion{Na}{i} absorbers toward 15 quasars. The columns give the name of the quasars, the coordinates, the
velocities in the LSR frame, the logarithm of the \ion{Ca}{ii} column densities,
the $b$-value for the \ion{Ca}{ii} lines, the logarithm of the \ion{Na}{i}
column densities, the $b$-value for the \ion{Na}{i} lines, the \ion{H}{i} column
densities, the $b$-value for the \ion{H}{i} lines and the possibly associated HVC/IVC complex. The $4\sigma$ and
$2\sigma$ detection limit (Section\,\ref{data}) for \ion{Ca}{ii}~$\lambda
3934.77$ and \ion{Ca}{ii}~$\lambda 3969.59$, respectively, corresponds to a
$\log (N_\mathrm{CaII}/\mathrm{cm}^{-2})$ detection limit of $\approx 11$. n/a
means that no Effelsberg data are available for these sight lines.}
\label{tab_20_quasars}
\small
\centering
\begin{tabular}{ccccccccccc}\hline\hline
\rule{0pt}{3ex}Quasar&$l$&$b$&$v_\mathrm{LSR}$&$\log N_\mathrm{CaII}$&$b_\mathrm{CaII}$&$\log N_\mathrm{NaI}$&$b_\mathrm{NaI}$& $\log N_\mathrm{HI}$&$b_\mathrm{HI}$&HVC/IVC\\

\rule{0pt}{3ex}&[$\degr$]&[$\degr$]&[km\,s$^{-1}$]&[$N$ in cm$^{-2}$]&[km\,s$^{-1}$]& [$N$ in cm$^{-2}$]&[km\,s$^{-1}$]&[$N$ in cm$^{-2}$]&[km\,s$^{-1}$]&complex\\[1ex]\hline
\rule{0pt}{3ex} QSO B1212+3326&173.1&80.1&$-$37&11.0&4&11.0&2&-&-&IV Arch?\\ 
&&&$-$49&11.5&2&10.5&2&19.7&18.1&IV Arch?\\
&&&$-$57&11.6&8&-&-&-&-&IV Arch?\\
&&&$-$76&11.0&4&-&-&-&-&IV Arch?\\\hline
QSO J1039-2719&270.0&27.0&184&11.3&5&-&-&-&-&WD\\
&&&$-11$&12.6&1&11.6&6&20.1&6.8&-\\\hline
J123437-075843&290.4&70.4&88&11.6&2&-&-&-&-&-\\
&&&74&11.9&7&-&-&-&-&-\\\hline
QSO B1331+170&348.5&75.8&$-$9&12.4&11&11.3&6&18.9&2.7&-\\
&&&$-$27&12.0&5&11.3&3&-&-&IV Spur\\\hline
J144653+011355&354.7&52.1&16&11.8&4&-&-&-&-&-\\
&&&$-$48&11.5&4&-&-&-&-&-\\\hline
QSO B2314-409&352.0&$-$66.3&$-$37&12.1&7&-&-&n/a&-&IV Spur\\
&&&$-$56&11.9&5&-&-&n/a&-&IV Spur\\\hline
QSO J1356-1101&327.7&48.7&$-$108&12.0&6&-&-&n/a&-&-\\\hline
QSO B0458-0203&201.5&$-$25.3&$-$7&12.1&4&12.1&2&19.3&2.6&AC shell\\\hline
QSO J0103+1316&127.3&$-$49.5&$-$351&11.7&4&-&-&n/a&-&-\\\hline
QSO J0153-4311&268.9&$-$69.6&124&12.4&6&-&-&n/a&-&MS\\\hline
QSO B0216+0803&156.9&$-$48.7&11&12.1&1&-&-&n/a&-&-\\\hline
QSO B2348-147&72.1&$-$71.2&91&11.4&1&-&-&n/a&-&MS\\\hline
QSO J0139-0824&156.2&$-$68.2&$-$100&12.2&2&-&-&n/a&-&MS\\\hline
QSO B0112-30&245.5&$-$84.0&$-$13&11.9&2&13.1&1&n/a&-&-\\\hline
QSO J0105-1846&144.5&$-$81.1&192&11.4&1&-&-&n/a&-&MS\\
&&&167&11.2&3&-&-&n/a&-&MS\\
&&&108&11.4&2&11.0&3&n/a&-&MS\\\hline


\end{tabular}
\end{table*}

The directions of several of the intermediate- and high-velocity absorbing systems as well as their velocities indicate a possible association with known large and extended HVC or IVC complexes, independent of whether they are detected in 21\,cm or not. Other sight lines, in contrast, do not appear to be associated with any known HVC or IVC complex. Obviously, optical absorption spectra allow us to trace the low neutral column density environment of the Galactic halo. The distribution of neutral or weakly ionised gas is more complex than indicated by \ion{H}{i} 21\,cm observations alone. 

Although the column densities have been measured with high accuracy, observations with high-resolution synthesis telescopes would be required to determine reliable Ca or Na abundances of the intermediate- and high-velocity gas. Another difficulty of Ca and Na measurements are the complex ionisation properties of calcium and sodium as well as dust depletion effects which introduce uncertainties in elemental abundance studies of these species \citep[see][]{wakkermathis00}. Note, that ionised material may represent a significant if not dominant fraction of the total gas amount of these gaseous material. The ionisation stage and the depletion of an element depends on the physical conditions in the halo, which may vary from location to location \citep{sembachsavage96}. Another critical aspect is that many of the sight lines have blending problems with high-redshift ($z > 2$) Ly$\alpha$ forest lines. The consequence is that a significant fraction of \ion{Ca}{ii} and \ion{Na}{i} high-velocity features may even remain unnoticed.  

\section{Statistical properties}\label{stat.properties} 

We performed a statistical analysis of the 55 \ion{Ca}{ii} and 20 \ion{Na}{i} intermediate and high velocity absorption components detected along 35 sight lines. In the following we turn our attention only to the \ion{Ca}{ii} absorption 
lines because there are only a few \ion{Na}{i} absorption components compared to a large number of \ion{Ca}{ii} detections. 

\begin{figure} 
\includegraphics[width=0.45\textwidth,bb=17 570 360 845,clip=]{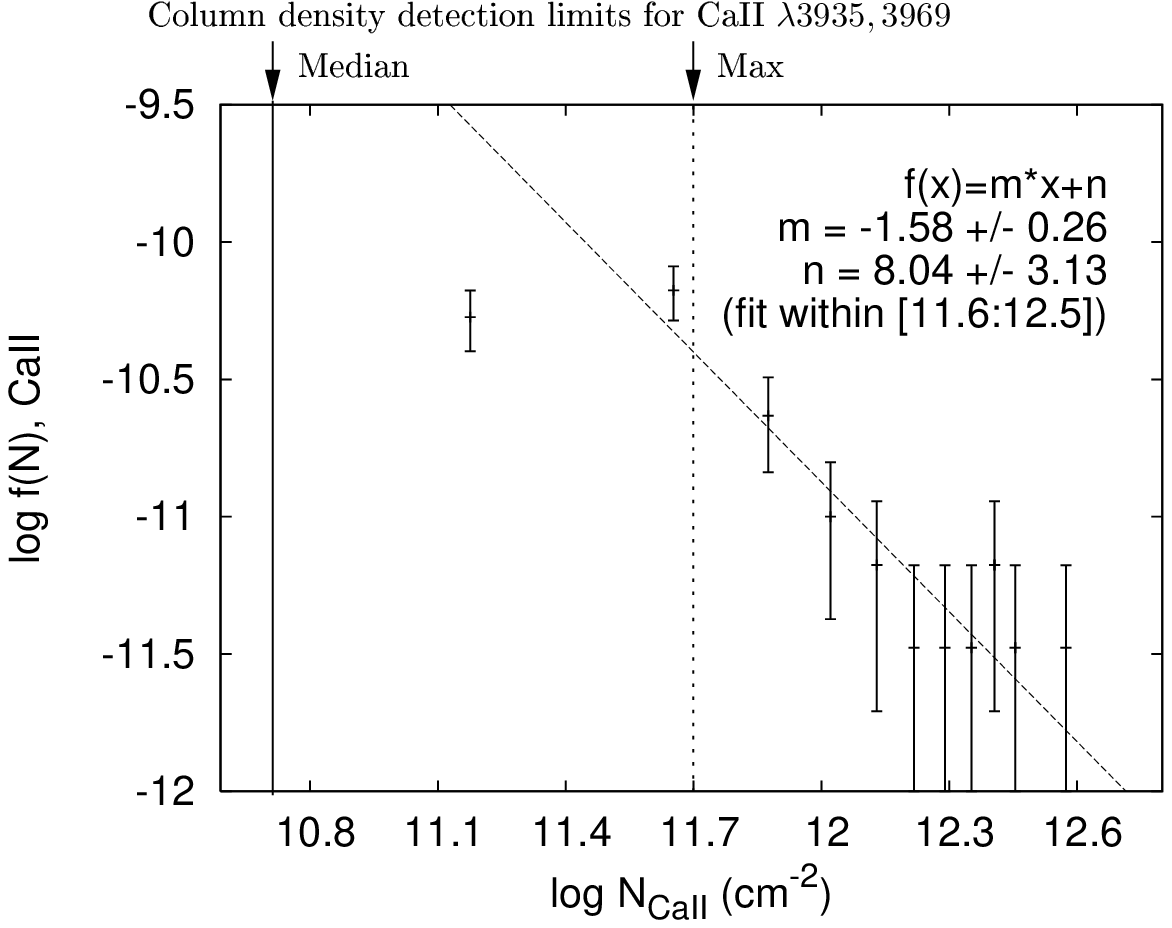}\hfill
\caption{The \ion{Ca}{ii} column density distribution $f(N)$ derived from UVES data. The dashed line represents a power-law fit $f(N)=CN^{\beta}$ with $\beta=-1.6 \pm 0.3$ and $\log C=8.0 \pm 3.1$ defined over the range of column densities from $\log (N_\mathrm{CaII}/\mathrm{cm}^{-2})=11.6$ to $\log (N_\mathrm{CaII}/\mathrm{cm}^{-2})=12.5$. The vertical solid line indicates the UVES $4 \sigma$ detection limit $\log (N_\mathrm{CaII}/\mathrm{cm}^{-2})=10.7$ for the median $S/N_\mathrm{r}$ and the dotted line represents the maximal detection limit $\log (N_\mathrm{CaII}^\mathrm{max}/\mathrm{cm}^{-2})=11.7$ for the minimal $S/N_\mathrm{r}$. The minimal detection limit $\log (N_\mathrm{CaII}^\mathrm{min}/\mathrm{cm}^{-2})=10.4$ for the maximal $S/N_\mathrm{r}$ is not plotted in this figure. The vertical errorbars are Poisson errors.}
\label{fig_num_vs_columndensity} 
\end{figure}

\begin{figure} 
\includegraphics[width=0.45\textwidth,bb=18 573 361 843,clip=]{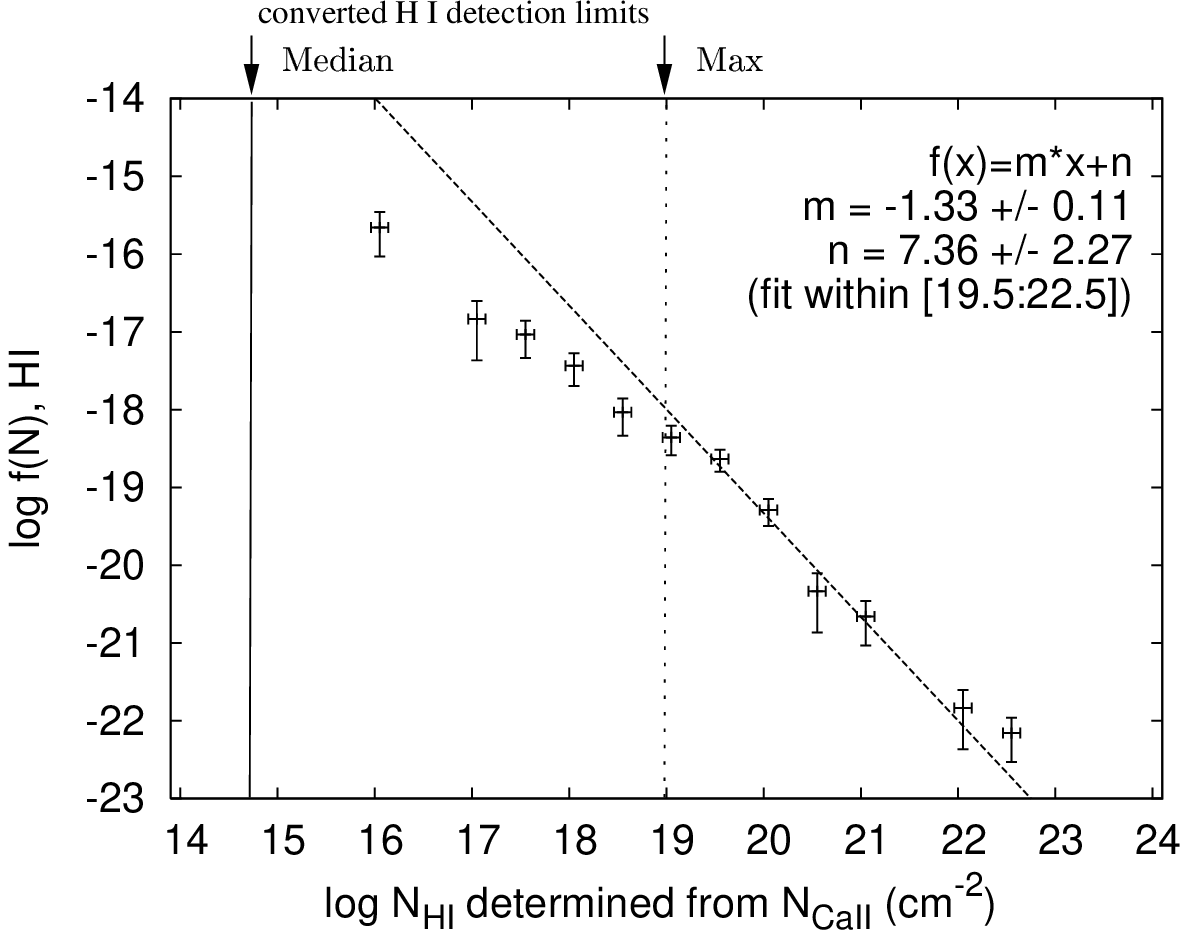}\hfill
\caption{The column density distribution function $f(N)$. The \ion{H}{i} column densities are calculated from the \ion{Ca}{ii} column densities detected with UVES with the help of the correlation between the abundance A(\ion{Ca}{ii}) and $N_\mathrm{HI}$ \citep{wakkermathis00}. The dashed line represents a power law fit $f(N)=C N^{\beta}$ with $\beta=-1.3 \pm 0.1$ and $\log C=7.4\pm2.3$ defined over the range of column densities from $\log (N_\mathrm{HI}/\mathrm{cm}^{-2})=19.5$ to $\log (N_\mathrm{HI}/\mathrm{cm}^{-2})=22.5$. The vertical and horizontal errorbars indicate Poisson and statistical (from Voigt profile fitting) errors, respectively. 
The vertical solid and dotted lines indicate the converted $4\,\sigma$ median and maximal UVES detection limit.} 
\label{fig_lognum_vs_logHIcolumndensity_UVES} 
\end{figure} 

\begin{figure} 
\includegraphics[width=0.45\textwidth,clip=]{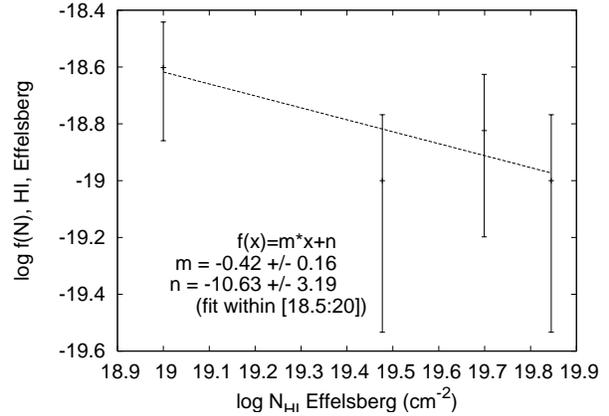}\hfill
\caption{The \ion{H}{i} column density distribution observed with Effelsberg. The dashed line represents a power law fit over the range of column densities ($\log (N_\mathrm{HI}/\mathrm{cm}^{-2})$) from 18.5 to 20. The vertical errorbars indicate Poisson errors.} 
\label{fig_lognum_vs_logHIcolumndensity_Eff} 
\end{figure}

\subsection{Column densities} 

Fig.\,\ref{fig_num_vs_columndensity} shows the \ion{Ca}{ii} column density distribution (CDD) function, $f(N)$, derived from the UVES data. Following \citet{churchillvogtcharlton03}, we define

\begin{equation} 
f(N)=\frac{m}{\Delta N}, 
\end{equation}

where $m$ is the number of clouds in the column density range $[N,N+\Delta N]$. Integrating $f(N)$ over the observed column density range yields the total number of clouds, $M$, as
\begin{equation}
M = \int \limits_{N_\mathrm{min}}^{N_\mathrm{max}}\mathrm{d}N\,f(N)\,. 
\end{equation}

The distribution of the \ion{Ca}{ii} column densities for $\log N_\mathrm{CaII}>11.6$ follows a power law $f(N)=CN^{\beta}$ with $\beta=-1.6 \pm 0.3$ and $\log C=8.0 \pm 3.1$. The vertical and horizontal errorbars indicate Poisson and statistical (from Voigt profile-fitting) errors, respectively. The vertical solid line indicates the UVES $4 \sigma$ detection limit $\log (N_\mathrm{CaII}/\mathrm{cm}^{-2})=10.7$ for the median $S/N_\mathrm{r}$ and the dotted line represents the maximal detection limit $\log (N_\mathrm{CaII}^\mathrm{max}/\mathrm{cm}^{-2})=11.7$ for the minimal $S/N_\mathrm{r}$. The minimal detection limit $\log (N_\mathrm{CaII}^\mathrm{min}/\mathrm{cm}^{-2})=10.4$ for the maximal $S/N_\mathrm{r}$ is not plotted in Fig.\,\ref{fig_num_vs_columndensity}. Note, that only the statistical noise is taken into account. The flattening of the distribution towards lower column densities therefore is caused by selection effects and other systematic errors (e.g., blending effects).

To better understand the relation between absorption-selected IVCs and HVCs from UVES and and the 21\,cm halo clouds seen with Effelsberg it would be very interesting to compare the column density distributions of these two data sets.
It is known that \ion{Ca}{ii} qualitatively traces neutral gas in the interstellar medium (ISM). However, since \ion{Ca}{ii} is not the dominant ionisation stage of calcium in the diffuse ISM and calcium is depleted into dust grains the conversion between \ion{Ca}{ii} column densities and \ion{H}{i} column densities is afflicted with large systematic uncertainties and thus has to be used with caution. Yet, it is an observational fact that there exists such relation between the column densities of these two ions in the ISM as shown by \citep{wakkermathis00}. These authors find a correlation between the abundance A(\ion{Ca}{ii}) and $N_\mathrm{HI}$ from 21\,cm data in the form 

\begin{equation} 
\log \left(N_\mathrm{HI}/\mathrm{cm}^{-2}\right) =  \frac{\log \left(N_\mathrm{CaII}/\mathrm{cm}^{-2}\right)-7.45}{0.22}
\label{eq_conversion}. 
\end{equation}

Although the scatter in this relation is substantial (see \citet{wakkermathis00}; their Fig.\,1) Eq.(5) allows us to roughly estimate \ion{H}{i} column densities from our measured \ion{Ca}{ii} column densities and to compare the column density distributions from absorption and emission. 
Note that at the current state we are not able to calculate a relation between \ion{Ca}{ii} and \ion{H}{i} from our 21\,cm data as we would have only ten data points available. 

Fig.\,\ref{fig_lognum_vs_logHIcolumndensity_UVES} shows the converted \ion{H}{i} column density distribution function $f(N)$. 
The dashed line represents a power law fit $f(N)=CN^\beta$, with $\beta=-1.3 \pm 0.1$ and $\log C=7.4\pm2.3$ defined over the range of column densities from $\log (N_\mathrm{HI}/\mathrm{cm}^{-2})=19.5$ to $\log (N_\mathrm{HI}/ \mathrm{cm}^{-2}) =22.5$. Fig.\,\ref{fig_lognum_vs_logHIcolumndensity_Eff} shows $f(N)$ for \ion{H}{i} observed with Effelsberg in the same directions where we find \ion{Ca}{ii}/\ion{Na}{i} absorption lines with UVES. The dashed line represents a power law fit with $N_\mathrm{Eff}^{-0.4}$ defined over the range of column densities from $\log (N_\mathrm{HI}/\mathrm{cm}^{-2})=18.5$ to $\log (N_\mathrm{HI}/\mathrm{cm}^{-2})=20$. Note, that we have only 14 absorption features along 13 lines of sight through the Galactic halo observed with the Effelsberg telescope. Therefore, the significance of the fit result is very poor. Furthermore, there is a bias in the Effelsberg data, because we only observed sight lines where \ion{Ca}{ii} absorption was detected.

Apparently, the slopes for the two different data sets deviate from each other. The most likely reason for this difference is
that the above-used \ion{Ca}{ii}-to-\ion{H}{i} column density conversion indeed is not appropriate for our absorbers because of the substantial uncertainties 
regarding ionisation, dust depletion and beam smearing effects. 
Note that the slope of our \ion{H}{i} column density distribution (Fig.\,\ref{fig_lognum_vs_logHIcolumndensity_UVES}) shows a slight flattening at low column densities ($\log (N_\mathrm{HI}/\mathrm{cm}^{-2}) < 19$). This is most likely due to incompleteness caused by the detection limit of the UVES instrument. Another reason for the change in slope at small column densities could be the fact that we take both IVC and HVC gas as one sample into account for the statistical investigation. By using the correlation  between $A(\ion{Ca}{ii})$ and $N_\mathrm{HI}$ \citep{wakkermathis00} we implicitly assume that IVCs and HVCs have the same intrinsic calcium abundance and similar dust-depletion properties, which is unrealistic because both populations probably have different origins \citep[Galactic versus extragalactic; see][
for a review] {richterp06}. We will be able to discuss these effects in more detail when we have a larger sample of \ion{Ca}{ii} absorbers, so that we can separate intermediate- and high-velocity \ion{Ca}{ii} absorbers and treat them as different samples to get adequate statistical results for both populations.

\subsection{Deviation velocities}\label{Deviation velocities}

\begin{figure} 
\includegraphics[width=0.5\textwidth, bb=70 65 420 275, clip=]{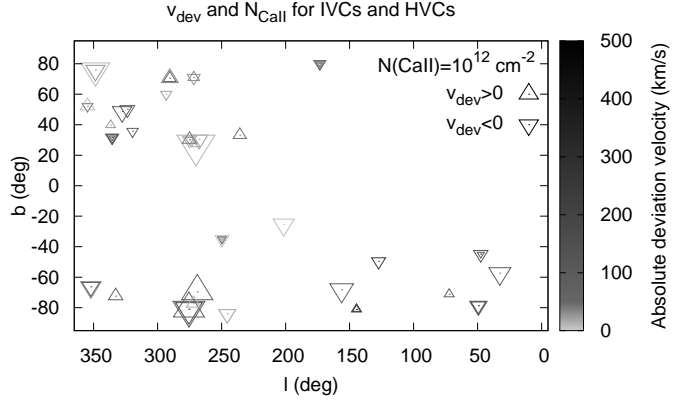}
\caption{Distribution of deviation velocities of the observed intermediate- and high-velocity absorbing systems versus galactic longitude and latitude. The dimensions of the triangles are proportional to the \ion{Ca}{ii} column densities observed with the UVES instrument. In the upper right corner of the diagram an exemplary \ion{Ca}{ii} column density of $N_\mathrm{CaII}=1 \times 10^{12}$\,cm$^{-2}$ is shown.} \label{fig_deviation_velocity_HVC_IVC} 
\end{figure}

Fig.\,\ref{fig_deviation_velocity_HVC_IVC} shows the deviation velocity \citep{wakker91} of all observed intermediate- and high-velocity absorbers versus galactic longitude and latitude.  The size of the triangles in Fig.\,\ref{fig_deviation_velocity_HVC_IVC} is proportional to the \ion{Ca}{ii} column densities observed with UVES. Obviously, there is no systematic trend in the velocity distribution. Note that we have no \ion{Ca}{ii}/\ion{Na}{i} UVES data for the region $b>0^\circ$ and $l<200^\circ$. Therefore, our $v_\mathrm{dev}$ distribution is incomplete. This fact also becomes apparent in Fig.\,\ref{fig_hvc_allskymap}.
We expect additional HST/STIS and KECK data for \ion{Ca}{ii}, \ion{Na}{i} and other species to cover this region of the sky (Richter et al. 2008, in preparation).  

\begin{figure} 
\includegraphics[width=0.45\textwidth,bb=50 45 410 302,clip=]{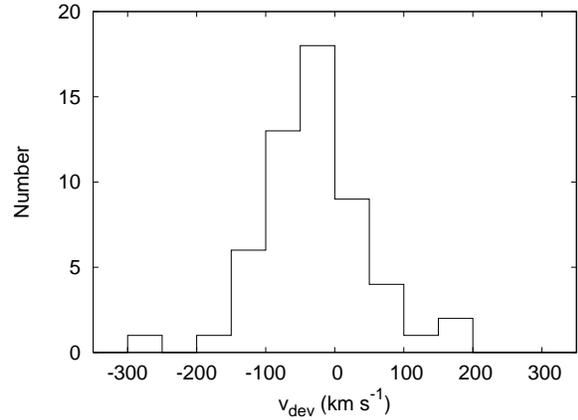}
\caption{Number of sight lines with intermediate- or high-velocity \ion{Ca}{ii} absorption lines versus deviation velocity.} 
\label{fig_numvsvdev_IVC_HVC_zusammen} 
\end{figure}

Fig.\,\ref{fig_numvsvdev_IVC_HVC_zusammen} shows
the number of sight lines with intermediate or high velocity
\ion{Ca}{ii}/\ion{Na}{i} absorption lines plotted against deviation velocity. There is an accumulation at low deviation velocities ($|v_\mathrm{dev}|<100\,\mathrm{km\,s}^{-1}$). In addition, there appears to be a slight excess at
negative $v_\mathrm{dev}$. It is unlikely that this is due to a selection effect
caused by our inhomogeneous all-sky sample. In the missing northern part of the sky most of the HVCs have negative LSR velocities due to the Galactic rotation. Therefore we also expect negative deviation velocities for this region of the sky. The slight excess for negative $v_\mathrm{dev}$ in our dataset is most likely due to systematic effects in connection with homogeneous structures infalling onto the Galactic disk. 

Note that there is one sight line (QSO\,J0103$+$1316) with a very high deviation velocity of about $v_\mathrm{dev}\approx -260\, \mathrm{km\,s}^{-1}$. This sight line passes the region between the Anti-Centre Complex and the Magellanic Stream, where high negative LSR velocities are widely observed. 

We also have searched for a possible correlation between the \ion{Ca}{ii} column densities and the deviation velocities. However, no such correlation is visible at this point. With 150 additional sight lines being available soon, we will be able to improve our statistical results.

\subsection{Doppler $b$-parameters and component structure} 

\begin{figure} 
\includegraphics[width=0.45\textwidth,clip]{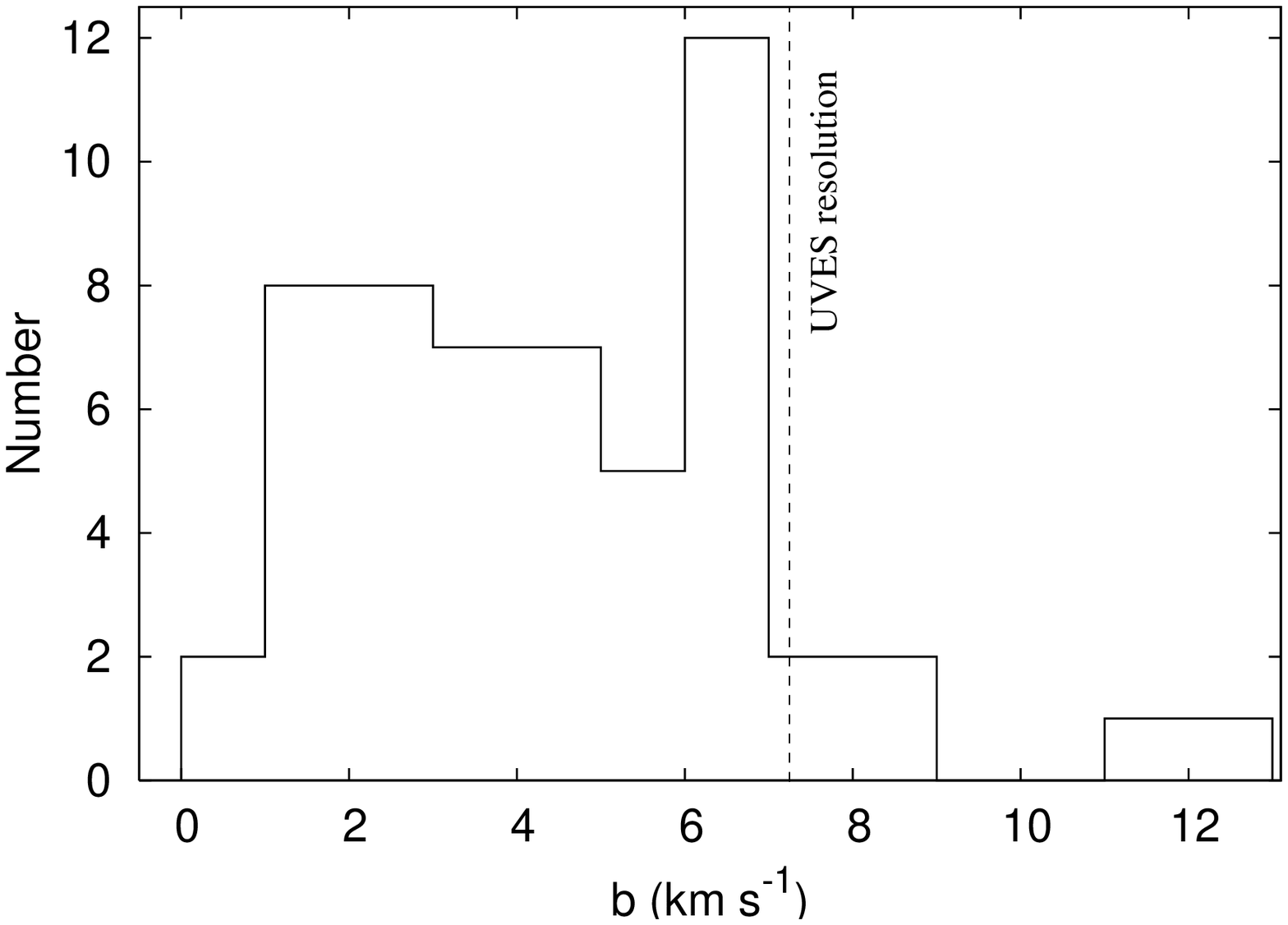}\hfill
\caption{Number of intermediate- and high-velocity \ion{Ca}{ii} absorption lines versus Doppler-parameter ($b$-value). The dashed vertical line indicates the UVES velocity resolution of $\Delta v_\mathrm{UVES} \approx 7\,\mathrm{km\,s}^{-1}$.} 
\label{fig_bvalue} 
\end{figure} 

Fig.\,\ref{fig_bvalue} shows the distribution of Doppler-parameters ($b$-values) for the intermediate or high velocity \ion{Ca}{ii} absorbers. The dashed vertical line indicates the UVES velocity resolution of
$\Delta v_\mathrm{UVES} \approx 7\,\mathrm{km\,s}^{-1}$. Therefore, lines with
$b<7\,\mathrm{km\,s}^{-1}$ are not resolved with the UVES instrument. However, the
histogram indicates that the bulk of the \ion{Ca}{ii} lines have $b$ values smaller than $7\,\mathrm{km\,s}^{-1}$, and only a small fraction has $b>7\,\mathrm{km\,s}^{-1}$ (as securely
obtained by the simultaneous fit of the two \ion{Ca}{ii} lines that have
different $f$ values). We can derive an upper temperature limit if we assume that the line width is dominated by thermal Doppler broadening.
\begin{equation}
T < \frac{b_\mathrm{therm}^2 A}{f} 
\end{equation}
%
%
In this expression $b$ is the Doppler parameter, $A$ is the atomic weight and $f=2 \cdot 10^{-6} \cdot \mathrm{k}/u$ with $\mathrm{k}$ the Boltzmann constant and $u$ the atomic mass unit. Assuming pure thermal line broadening a $b$-value of $7\,\mathrm{km\,s}^{-1}$ and $A_\mathrm{CaII}=40$ corresponds to a temperature of $T\approx 1.2 \cdot 10^5\,K$.

Since \ion{Ca}{ii} absorption traces mostly neutral gas, it is clear, however, that the $b$-value is dominated by turbulent gas motions rather than by the temperature of the gas. This is shown also by the much lower temperatures estimates from the measured 21\,cm line widths (see Section\,3).
The fact that the bulk of the \ion{Ca}{ii} lines have $b$ values of $< 7\,\mathrm{km\,s}^{-1}$ indicates that turbulence and bulk motions in the gas are relatively small, suggesting that the absorbers are spatially confined in relatively small clumps on pc-scales. \citet{sembach03} used FUSE data to study highly ionised (\ion{O}{vi}) high-velocity absorbing systems along sight lines through the Galactic halo in the directions of 100 extragalactic objects. Compared to our absorbers these highly ionised systems have much higher $b$-values in the range of $16\ldots72$\,km\,s$^{-1}$, suggesting that high-ion absorbers live in much more extended regions (kpc-scales) than the \ion{Ca}{ii} absorbers.

Systems that show \ion{Na}{i} absorption possibly are even more confined, as the presence of \ion{Na}{i} requires relatively high gas densities ($n_H > 0.1$ cm$^{-3}$) to achieve a detectable level of neutral sodium, which has an ionisation potential of $<10$~eV. For a constant thermal gas pressure, on the other hand, high densities imply low temperatures of the gas.

This idea is supported by high-resolution ($R\sim68000$) observations by \citet{sembachdankssavage93} and \citet{sembachdanks94S}. They obtained a sample of spectra in the direction of 55 stars at distances above 2\,kpc, containing 231 \ion{Na}{i} and 312 \ion{Ca}{ii} components. 10\% of the total \ion{Ca}{ii} column densities were found at forbidden velocities, while in case of \ion{Na}{i} this is true only for a small fraction of few percent. The absorbers show a two-component temperature distribution of cold ($\sim100\,$K) and warm ($\sim1000$ to $5000\,$K) gas in diffuse clouds and intercloud medium. \ion{Ca}{ii} traces both media, while \ion{Na}{i} was detected preferentially in the cloudy medium.

\begin{table}
\caption{Number of IVC/HVC \ion{Ca}{ii} absorption components observed with UVES and their absolute and percental values. The total number of sight lines which show \ion{Ca}{ii} absorption lines is 31. The last two columns show the data analysed by \citet{prochteretal06} for comparison. Their total number of sight lines is 25.}
\label{tab_7_components}
\small
\centering
\begin{tabular}{ccccc}\hline\hline
\rule{0pt}{3ex}Number of &Sight lines with &  & Sight lines with & \\
components &\ion{Ca}{ii} absorption & [\%] & \ion{Mg}{ii} absorption & [\%]\\\hline
\rule{0pt}{3ex}1&17&55&5&20\\
 2&7&23&13&52\\
 3&4&13&2&8\\
 4&3&1&3&12\\
 5&0&0&1&4\\
 6&0&0&0&0\\
 7&0&0&1&4\\\hline

\end{tabular}
\end{table}

Note that the presence of relatively cold, sub-pc and AU-scale structures in the Galactic halo gas has been proven by other observations.
One sight line in the direction of PKS\,1448$-$232 with distinct \ion{Ca}{ii}/\ion{Na}{i} absorption and corresponding \ion{H}{i} emission lines was recently observed by \citet{richterwestmeierbruens05} with the VLA. At an angular resolution of about $2'$\,HPBW they detected several  cold ($T<1000$\,K) clumps of neutral gas with fairly low peak column densities of $N_{\rm HI} \approx 7 \cdot 10^{18} \, {\rm cm}^{-2}$. In addition, ultraviolet measurements of molecular hydrogen in Galactic extraplanar clouds have shown that small, dense gaseous clumps at sub-pc scale are widespread in the lower halo, in particular in the intermediate-velocity clouds \citep{richterdeboeretal99, richtersavagesembachetal03, richterwakkersavagesembach03}.

To check whether a single or a multi-component absorption structure is
typical for the extraplanar \ion{Ca}{ii} absorbers we have created a table containing the
number of IVC/HVC \ion{Ca}{ii} absorption components (Table\,\ref{tab_7_components}). Obviously, absorption systems with a single or a double absorption-component structure are more common ($77\%$) than systems with more than two absorption components ($16\%$).

\section{Discussion of physical properties and origin of the absorbers}\label{disscussion} 

The fact that among 103~random lines of sight through the halo we observe
35 optical extraplanar absorption systems with at least one intermediate- or high-velocity component shows that the Milky Way halo contains a large number of neutral gas structures that give rise to intermediate- and high-velocity \ion{Ca}{ii} and \ion{Na}{i} absorption. 

\subsection{Area filling factors}

In the following we want to determine the area filling factors of our absorbers observed with UVES and Effelsberg. The area filling factor which we define is the fraction of sight lines which show \ion{Ca}{ii} absorption or \ion{H}{i} absorption to the total number of observed sight lines. We get a total \ion{Ca}{ii} area filling factor of about $30\%$ and a significantly larger \ion{H}{i} area filling factor for the Effelsberg data of about $68\%$. Both factors are defined for the same lower \ion{H}{i} column density regime as described above.
The discrepancy between the total absorption and emission area filling factors could possibly be explained by the effect that the neutral gas structures that give rise to the \ion{Ca}{ii} absorption only partly fill the $9\arcmin$ beam of the Effelsberg telescope. This could be an indication that the angular extent of the absorbers is small. The beam sizes of the different instruments (UVES, Effelsberg, LAB) also must have an effect on the column density distributions. Especially possible small-scale, high column density structures get affected by beam smearing, leading to a bias towards smaller column densities. The slope of the column-density distribution from 21\,cm data therefore is expected to steepen with stronger beam-smearing. Thus, the different slopes of the \ion{H}{i} column-density distributions from the UVES data and the Effelsberg data can be explained naturally by the limited beam size of the radio observations. 

We also determined the area filling factor for our intermediate- and high-velocity \ion{Ca}{ii} absorbers separately. We distinguish between IVCs and HVCs by using the deviation velocity of the clouds as a separation criterion (see Section\,\ref{Deviation velocities}). The \ion{Ca}{ii} area filling factor is about $16\%$ for the intermediate-velocity and about $17\%$ for the high-velocity gas for converted \ion{H}{i} column densities above $\log(N_\mathrm{HI}/\mathrm{cm}^{-2})\approx16$.
Using our Effelsberg observations, we calculate an \ion{H}{i} area filling factor of about $37\%$ (IVCs) and $16\%$ (HVCs) for densities above $\log(N_\mathrm{HI}/\mathrm{cm}^{-2})\approx18$.

The analysed UVES and Effelsberg spectra show that for several cases the
\ion{Ca}{ii} and \ion{Na}{i} absorbers have no intermediate- or high-velocity counterpart in the
21\,cm data obtained with Effelsberg. This implies that the \ion{H}{i} column
densities of the \ion{Ca}{ii} and \ion{Na}{i} absorption components at these
velocities fall below the detection limit of the Effelsberg 21\,cm observations
and also of current large surveys like the LAB survey. Another possibility for
this non-detection is a small diameter of the clouds, so that beam-smearing effects will again become important. 

\subsection{Possible association with known IVC or HVC complexes}
The directions of 19 of the intermediate- and high-velocity clouds as well as their velocities indicate a possible association with known large and extended IVC or HVC complexes (Tables\,\ref{tab_27_quasars} and \ref{tab_20_quasars}). 
Since we do not know the distances to the several absorbing systems the only criteria for the determination of a possible association with IVC/HVC complexes are the positions and the velocities.

Eight absorbing systems, and therefore the majority, are associated with the extensive Magellanic Stream (MS). In all eight cases the sight lines passes only the outer regions of the MS. In the case of QSO\,J0139$-$0824 and QSO\,J0105$-$1846 it is not clear if they are possibly parts of the MS or the Anti-Center shell (AC shell) because they are located in the intermediate range. Only for the sight line towards QSO~J0003$-$2323 we have corresponding \ion{H}{i} observations obtained with the 100-m telescope at Effelsberg.

Four systems are most likely associated with the intermediate-velocity Spur (IV-Spur). The sight lines toward QSO\,J1211$+$1030 and QSO\,B1331+170 pass the inner, denser part of the IV-Spur whereas the sight lines in the direction of QSO\,B0952$+$179 and QSO\,B2314$-$409 are located in the outer, clumpy structures of the complex. For three of these sight lines we have additional \ion{H}{i} observations but only in one case (QSO\,J1211$+$1030) we found a corresponding emission line.

The position and the velocity of the absorbing systems in the direction of QSO\,B1212$+$3326 indicate that these systems are possibly part of the inner region of the intermediate-velocity Arch (IV-Arch). For these sight line we have obtained additional \ion{H}{i} data. Only one component has a counterpart in \ion{H}{i}.

One absorption component in the direction of QSO\,J1039$-$2719 is most likely associated with the HVC complex WD. We re-observed this sight line with the 100-m telescope at Effelsberg but we find no corresponding \ion{H}{i} emission lines.

The sight line towards QSO\,B0458$-$0203 passes the very outer regions of the Anti-Center shell. The \ion{H}{i} data reveal a corresponding emission line.

The velocity and the coordinates of the absorption component with the high positive LSR velocity in the direction of QSO\,B1101$-$26 is possibly associated with the Leading Arm (LA) of the Magellanic System. The sight line passes the outer region of the LA. The Effelsberg data show no corresponding \ion{H}{i} emission line.

The absorbing systems towards QSO\,1448$-$232 are possibly associated with the HVC complex L \citep{richterwestmeierbruens05}.

The two absorbing components in the direction of QSO\,J2155$-$0922 are possibly part of the HVC complex GCN. The Effelsberg data show no corresponding \ion{H}{i} emission line.

\subsection{Comparison with \ion{Mg}{ii} systems and column density distribution functions}

It is evident that low-column density intermediate- and high-velocity \ion{Ca}{ii} absorbers have a large area filling factor because of the high \ion{Ca}{ii} and \ion{Na}{i} detection rate. If these absorbers represent a typical phenomenon of spiral galaxies like our Milky Way or M31 and if they are located in the halo of these galaxies, they should produce \ion{H}{i} Lyman-Limit and \ion{Mg}{ii} absorption along sight lines which pass through the halo gas of other galaxies. There are two different populations of \ion{Mg}{ii} absorbers, the weak ($W_\mathrm{\lambda 2796} \leq0.3\,\AA{}$) and the strong ($W_\mathrm{\lambda 2796}>0.3\,\AA{}$) \ion{Mg}{ii} systems. While the weak \ion{Mg}{ii} absorbers are thought to be located in the outskirts of galaxies \citep[e.g.,][]{churchill99}, the strong \ion{Mg}{ii} systems are most likely located in the halo \citep[e.g.,][]{petitjeanbergeron90, charltonandchurchill98, dingcharltonchurchillpalma03} or even in the discs of galaxies \citep[Damped Lyman Alpha systems, DLAs,][]{raoetal06}, and they possibly represent the analogues of IVCs/HVCs \citep[e.g.,][]{savageetal00}. In fact, many of the properties of the high-velocity \ion{Ca}{ii} and \ion{Na}{i} absorption lines resemble those of the strong \ion{Mg}{ii} absorption line systems observed in the circumgalactic environment of other galaxies 
\citep{dingetal03, boucheetal06, prochteretal06}. The high-velocity \ion{Ca}{ii} and \ion{Na}{i} absorbers thus may possibly represent the Galactic counterparts of strong \ion{Mg}{ii} systems at low redshift. 

\begin{figure*} 
\centering 
\includegraphics[width=0.9\textwidth,bb=90 128 505 715,clip=]{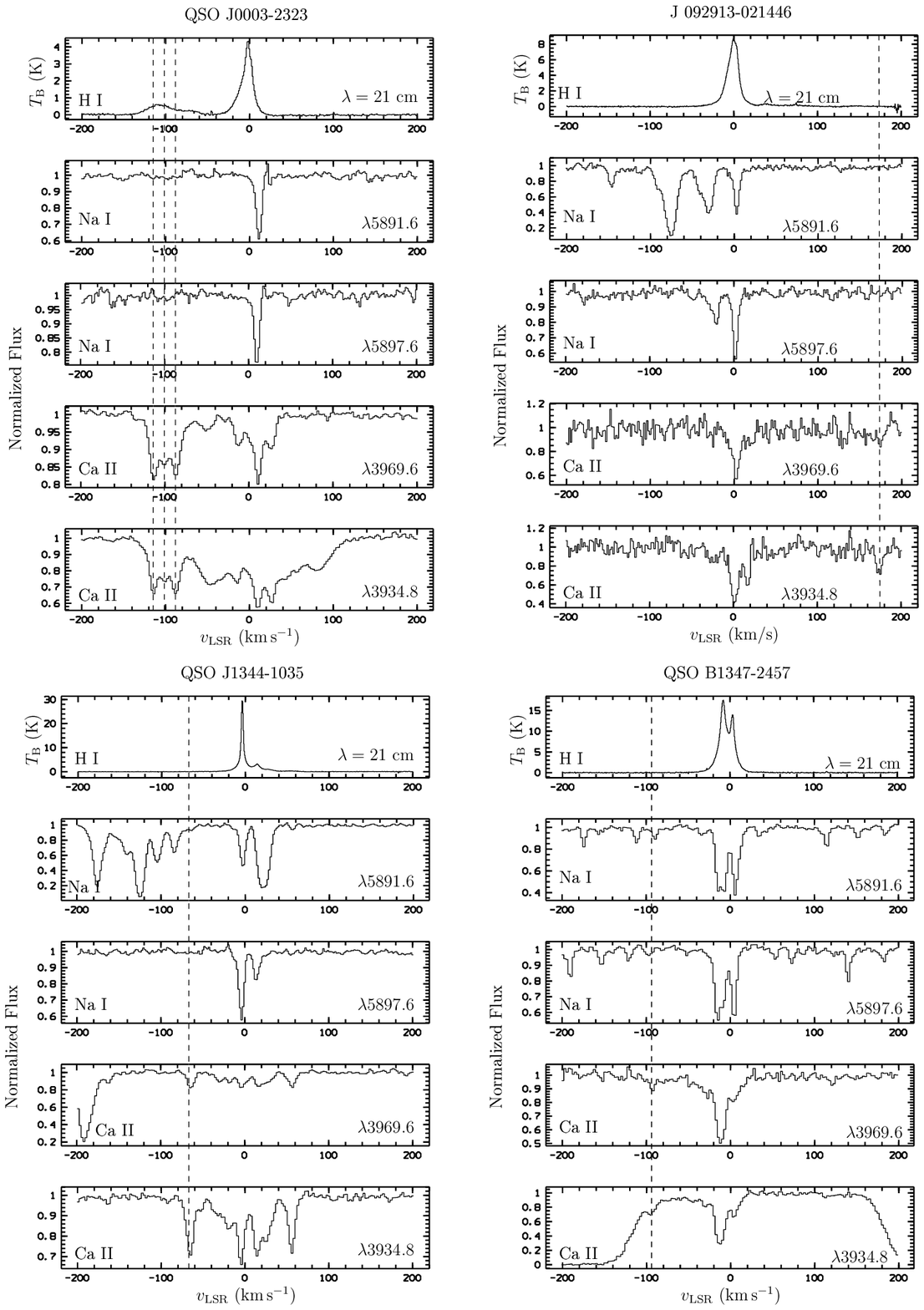}
\caption{\ion{Ca}{ii} and \ion{Na}{i} absorption and \ion{H}{i} emission spectra in the direction of the quasars QSO\,J0003$-$2323, J092913$-$021446, QSO\,J1344$-$1035, and QSO\,B1347$-$2457 obtained with UVES and the Effelsberg 100-m telescope, respectively. The absorption and corresponding emission lines are indicated by dashed lines.}
\label{fig_he0001-2340_he0151_he1341_J092913} 
\end{figure*}

\begin{figure*} 
\centering 
\includegraphics[width=0.9\textwidth,bb=90 128 505 715,clip=]{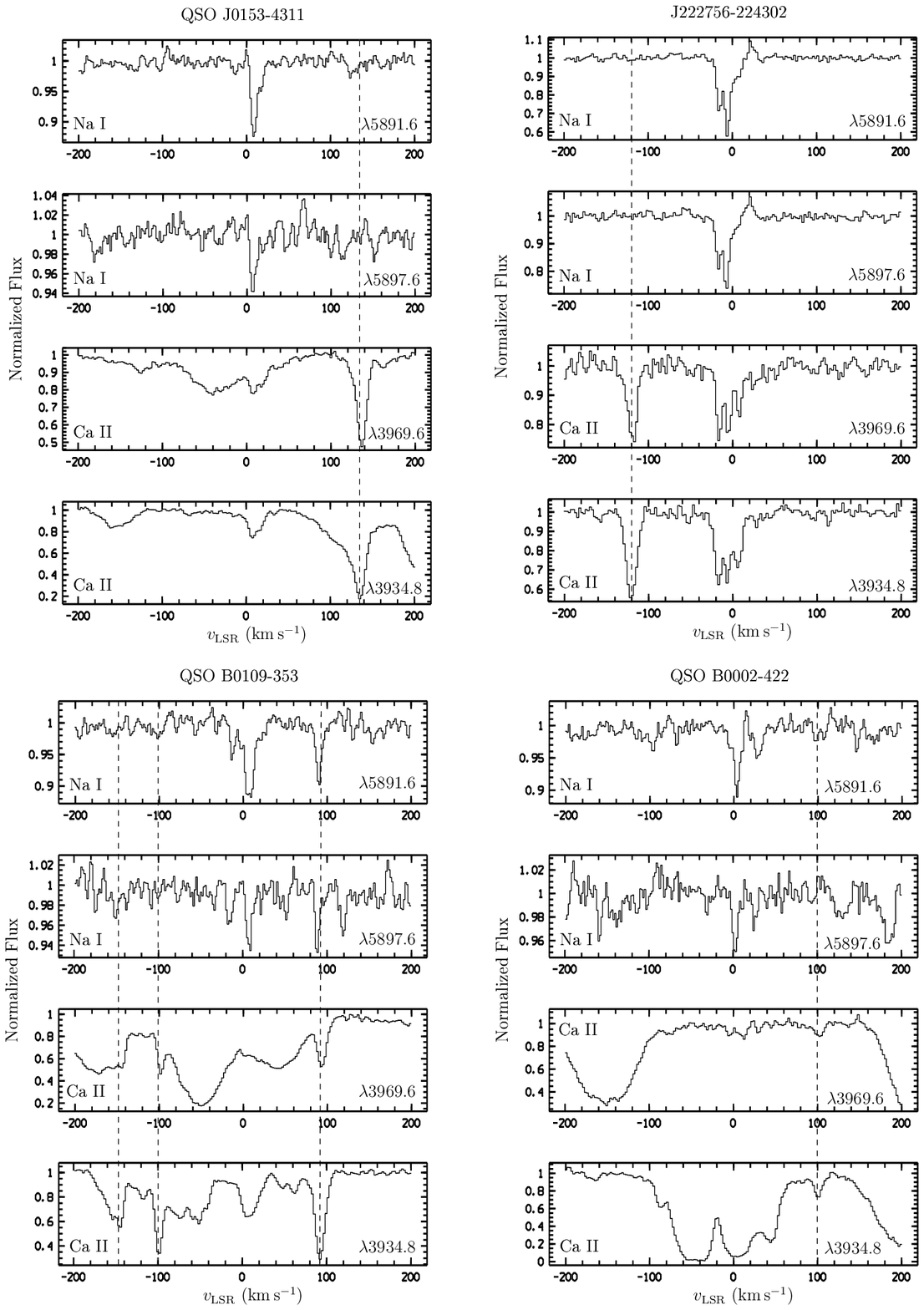}
\caption{\ion{Ca}{ii} and \ion{Na}{i} absorption spectra of QSO\,J0153$-$4311, J222756$-$224302, QSO\,B0109$-$353, and QSO\,B0002$-$422 obtained with UVES. For these sight lines no Effelsberg data are available.} 
\label{fig_he0151_J222756_q0109_q0002} 
\end{figure*}

\begin{figure*} 
\centering 
\includegraphics[width=0.9\textwidth,bb=90 128 505 715,clip=]{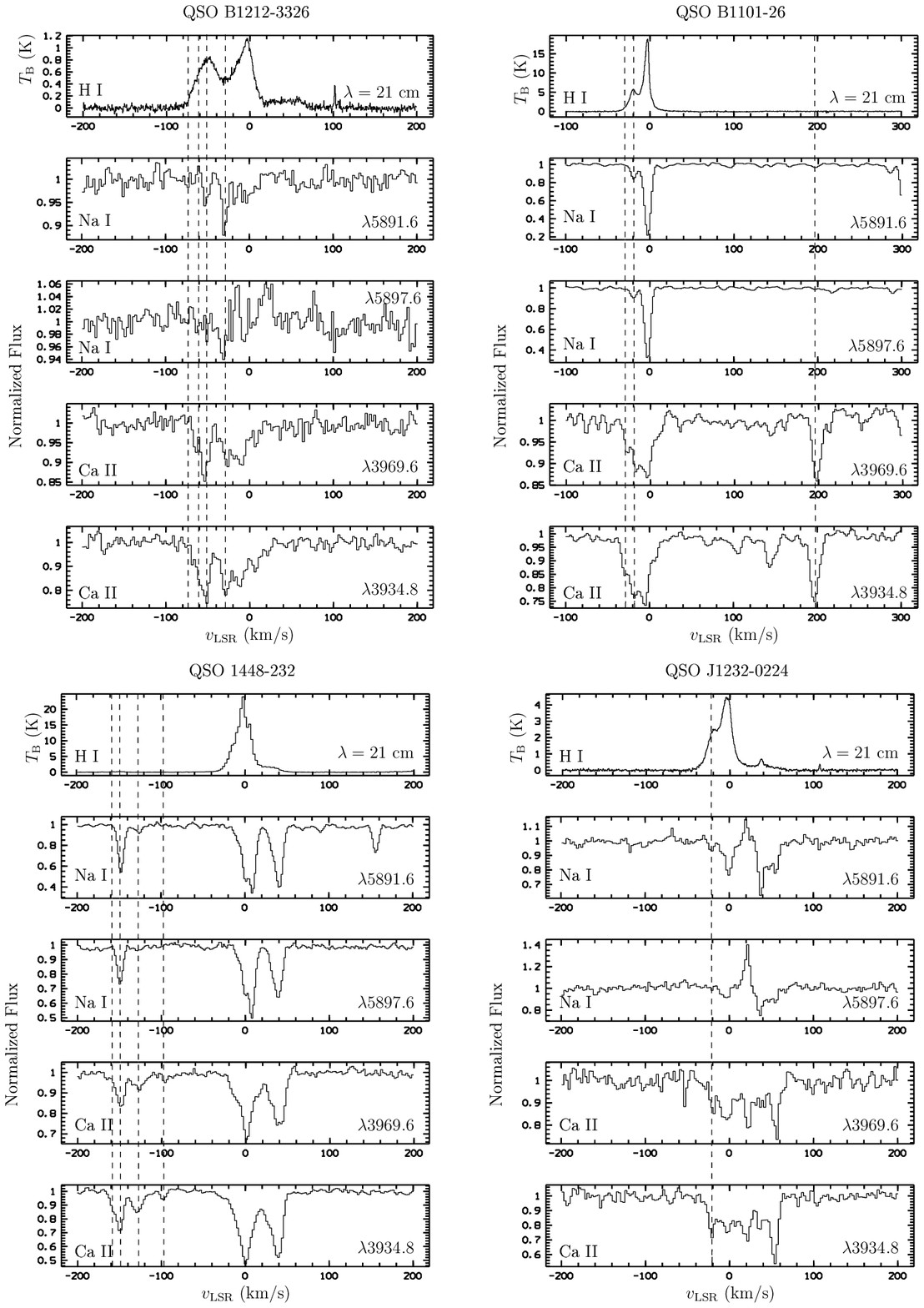}
\caption{\ion{Ca}{ii} and \ion{Na}{i} absorption and corresponding \ion{H}{i} emission spectra of QSO\,B1212$+$3326, QSO\,B1101$-$26, QSO\,1448$-$232, and QSO\,J1232-0224 obtained with UVES and the Effelsberg 100-m telescope, respectively.} 
\label{fig_QSOB1212_QSOB1101_PKS1448-232_LBQS1229-0207} 
\end{figure*}

In Section\,\ref{stat.properties} we have shown that absorption systems at intermediate and high velocities with a single
or a double absorption component structure are more common than systems with more than two absorption components. The analysis of strong \ion{Mg}{ii} absorption systems by \citet{prochteretal06} shows that there is a tendency for two or more \ion{Mg}{ii} absorption lines (Table\,\ref{tab_7_components}). The more complex absorption structure for \ion{Mg}{ii} systems can be explained by longer sight lines through the halo of the host galaxy compared to the typically shorter sight lines through the halo of the Milky Way. Our statistical analysis shows that there is no obvious difference in the observed parameters of single-component and 
multi-component absorption systems. As described by \citet{dingcharltonchurchill05}, multiple cloud components in absorption systems can be divided into two sub-classes. The ``kinematically spread'' absorbers show one or more dominant \ion{Ca}{ii} absorbers and several weaker
ones spread over a wide velocity range, as seen towards QSO 0109-3518 and QSO B1101-26 (Figures\,\ref{fig_he0151_J222756_q0109_q0002} and \ref{fig_QSOB1212_QSOB1101_PKS1448-232_LBQS1229-0207}, online version). 
The ``kinematically compact'' subclass is characterised by multiple absorption components with comparable equivalent widths blended together and spread over less than about $100\,\mathrm{km\,s}^{-1}$. An example of the latter sub-class is the absorber towards QSO J0003-2323, as displayed in Figure\,\ref{fig_he0001-2340_he0151_he1341_J092913}.

Important information about the distribution of neutral and weakly ionised extraplanar gas in the Milky Way is provided by column density distribution functions $f(N)$ of \ion{Ca}{ii} and \ion{H}{i} that we have derived in Section\,\ref{stat.properties}. We now can compare the properties of these distribution functions with results from QSO absorption-line studies and extragalactic \ion{H}{i} surveys. The slope of the \ion{Ca}{ii} column density distribution of Milky Way absorbers turns out to be $\beta=-1.6 \pm 0.3$. This is identical to the slope of $\beta=-1.59 \pm 0.05$ derived for strong \ion{Mg}{ii} absorbers at intermediate reshifts ($z=0.4-1.2$), as presented by \citet{churchillvogtcharlton03}. The equality of the slopes suggests that (despite somewhat different ionisation and dust-depletion properties) \ion{Ca}{ii} and strong \ion{Mg}{ii} absorbers probe similar gaseous structures that are located in the environment of galaxies. 

Concerning neutral hydrogen, \citet{petitjean93} have investigated the \ion{H}{i} column density distribution function of intergalactic QSO absorption line systems at high redshift (mean redshift of $z \approx 2.8$) with the help of high spectral resolution data. Their \ion{H}{i} data span a range from $\log (N_\mathrm{HI}/\mathrm{cm}^{-2})=12$ to $\log (N_\mathrm{HI}/\mathrm{cm}^{-2})=22$. After they could show that a single power law with a slope of $\beta=-1.49$ provides only a poor fit to the data they divided the sample into two subsamples (lower and higher than $10^{16}$\,cm$^{-2}$). They found a slope of $\beta=-1.83$ for the $\log (N_\mathrm{HI}/\mathrm{cm}^{-2}) < 16$ sample and $\beta=-1.32$ for the $\log (N_\mathrm{HI}/\mathrm{cm}^{-2}) > 16$ sample. \citet{petitjean93} point out that the overall \ion{H}{i} distribution is more complex than had been thought and that even two power laws only poorly fit the data. In a later study, \citet{kimcarswelletal02} indeed show that the slope of the \ion{H}{i} column density distribution varies between $\sim -1.4$ and $\sim -2.0$, depending on the mean absorber redshift and the column-density interval used. Unfortunately, very little is known about the \ion{H}{i} absorber distribution in the for us interesting range $\log (N_\mathrm{HI}/\mathrm{cm}^{-2})=15\ldots19$ and $z=0\ldots0.5$ due to the lack of low-redshift QSO data in the UV band.

Although we have substantial uncertainties in the conversion of \ion{Ca}{ii} into \ion{H}{i} as discussed above, it is interesting that the slope of the \ion{H}{i} column density distribution of $\beta=-1.3 \pm 0.1$, as indirectly derived from our \ion{Ca}{ii} data for the $\log (N_\mathrm{HI}/\mathrm{cm}^{-2})$ range between 19 and 22, is in general agreement with the statistics of low- and high-redshift QSO absorption line data. This implies that a significant fraction of high-column density (log $N>16$) intervening QSO absorbers are closely related to galaxies in a way similar as the intermediate- and high-velocity \ion{Ca}{ii} absorbers are connected to the Milky Way. It appears that the column density distribution of extraplanar neutral gas structures (i.e., HVCs and their extragalactic analogues) is roughly universal at low and high redshift. This underlines the 
Overall importance of the processes that lead to the circulation of neutral gas in the environment of galaxies (e.g., fountain flows, gas accretion, tidal interactions) for the evolution of galaxies.

\section{Summary and outlook}\label{summary} 

We studied optical absorption of \ion{Ca}{ii} and \ion{Na}{i} (archival VLT/UVES data) and \ion{H}{i} 21\,cm emission (Effelsberg telescope) in the direction of quasars. The major results of our project are:
\begin{enumerate}
\item Intermediate- and high-velocity \ion{Ca}{ii}/\ion{Na}{i} absorption is found along 35 out of 103 studied lines of sight in the direction of quasars. In total we found 55 individual absorption components.
\item In some cases the \ion{Ca}{ii} absorption lines are associated with known intermediate- and high-velocity clouds, but in other cases the observed absorption has no 21\,cm counterpart.
\item The observed \ion{Ca}{ii} column density distribution follows a power-law $f(N)=CN^{\beta}$ with a slope of $\beta \approx -1.6$. 
\item This distribution is similar to the distribution found for intervening \ion{Mg}{ii} systems that trace the gaseous environment of other galaxies at low and high redshift. After having transformed the observed \ion{Ca}{ii} column densities into \ion{H}{i} column densities we find that the extraplanar absorbers trace neutral gas structures with logarithmic \ion{H}{i} column densities on the order of $\log (N_\mathrm{HI}/\mathrm{cm}^{-2}) \approx 18 \ldots 20$.
\item The physical and statistical properties of the \ion{Ca}{ii} absorbers suggest that these structures represent the local counterparts of strong ($W_\mathrm{\lambda 2796}> 0.3\,\AA{}$) \ion{Mg}{ii} absorbing systems that are frequently observed at low and high redshift in QSO spectra and that are believed to trace the environment of other galaxies. 
\item Most of the absorbers are characterised by single-component absorption with $b$ values of $<7\,\mathrm{km\,s}^{-1}$.
\end{enumerate}

For the future we are planning to extend our study of extraplanar \ion{Ca}{ii} absorbers using additional high-resolution data from UVES as well as KECK data. Using radio synthesis telescopes, we are further planning to re-observe several sight lines with \ion{Ca}{ii}/\ion{Na}{i} absorption and \ion{H}{i} emission. These high-resolution observations are crucial to search for related small-scale \ion{H}{i} structures that are possibly not resolved with the Effelsberg telescope \citep[see also][]{hoffmanetal04, richterwestmeierbruens05}. Finally, we are currently analysing several ultraviolet QSO  spectra from HST/STIS to measure weak metal line absorption in the Milky Way halo (in particular, absorption from neutral and weakly ionised species such as \ion{O}{i}, \ion{Si}{ii}, and others). This will allow us to determine metallicities of the low-column density gas in the Milky Way halo. Metal abundance is an important parameter that will help to clarify whether the extraplanar gaseous structures are of Galactic or extragalactic origin. Furthermore, the high-resolution \ion{H}{i} data in combination with the VLT/UVES  and STIS data will enable us to study the physical parameters of these low column density gaseous structures, such as temperature and density, to better understand the connection between the different gas phases in the Milky Way halo. As mentioned before, the properties of the \ion{Ca}{ii} and \ion{Na}{i} absorption lines resemble those of the \ion{Mg}{ii} absorption line systems observed in the environment of other galaxies. Studying the Galactic halo with the proposed strategy thus will provide important new insights into absorption line systems that are commonly observed in the halos of more distant galaxies.

\begin{acknowledgements} Thanks to Benjamin Winkel and Peter Ernie for their helping hands. N.B.B. and P.R. acknowledge support by the DFG through DFG Emmy-Noether grant Ri\,1124/3-1. T.W. was supported by the DFG through grant KE 757/4-2. Based on observations with the 100-m telescope of the MPIfR (Max-Planck-Institut f\"{u}r Radioastronomie) at Effelsberg. 
\end{acknowledgements} 
\bibliographystyle{aa} 
\bibliography{9067} 

\begin{thebibliography}{57}
\expandafter\ifx\csname natexlab\endcsname\relax\def\natexlab#1{#1}\fi

\bibitem[{{Adams}(1949)}]{adams49}
{Adams}, W.~S. 1949, \apj, 109, 354

\bibitem[{{Arnal} {et~al.}(2000){Arnal}, {Bajaja}, {Larrarte}, {Morras}, \&
  {P{\"o}ppel}}]{arnalbajajaetal00}
{Arnal}, E.~M., {Bajaja}, E., {Larrarte}, J.~J., {Morras}, R., \& {P{\"o}ppel},
  W.~G.~L. 2000, \aaps, 142, 35

\bibitem[{{Bajaja} {et~al.}(2005){Bajaja}, {Arnal}, {Larrarte}, {Morras},
  {P{\"o}ppel}, \& {Kalberla}}]{bajajaarnaletal05}
{Bajaja}, E., {Arnal}, E.~M., {Larrarte}, J.~J., {et~al.} 2005, \aap, 440, 767

\bibitem[{{Bergeron} {et~al.}(2004){Bergeron}, {Petitjean}, {Aracil}, {Pichon},
  {Scannapieco}, {Srianand}, {Boisse}, {Carswell}, {Chand}, {Cristiani},
  {Ferrara}, {Haehnelt}, {Hughes}, {Kim}, {Ledoux}, {Richter}, \&
  {Viel}}]{bergeronpetitjeanetal04}
{Bergeron}, J., {Petitjean}, P., {Aracil}, B., {et~al.} 2004, The Messenger,
  118, 40

\bibitem[{{Bouch{\'e}} {et~al.}(2006){Bouch{\'e}}, {Murphy}, {P{\'e}roux},
  {Csabai}, \& {Wild}}]{boucheetal06}
{Bouch{\'e}}, N., {Murphy}, M.~T., {P{\'e}roux}, C., {Csabai}, I., \& {Wild},
  V. 2006, \mnras, 813

\bibitem[{{Braun} \& {Burton}(1999)}]{braunburton99}
{Braun}, R. \& {Burton}, W.~B. 1999, \aap, 341, 437

\bibitem[{{Bregman}(1980)}]{bregman80}
{Bregman}, J.~N. 1980, \apj, 236, 577

\bibitem[{{Charlton} \& {Churchill}(1998)}]{charltonandchurchill98}
{Charlton}, J.~C. \& {Churchill}, C.~W. 1998, \apj, 499, 181

\bibitem[{{Charlton} {et~al.}(2000){Charlton}, {Churchill}, \&
  {Rigby}}]{charltonetal00}
{Charlton}, J.~C., {Churchill}, C.~W., \& {Rigby}, J.~R. 2000, \apj, 544, 702

\bibitem[{{Churchill} {et~al.}(1999){Churchill}, {Rigby}, {Charlton}, \&
  {Vogt}}]{churchill99}
{Churchill}, C.~W., {Rigby}, J.~R., {Charlton}, J.~C., \& {Vogt}, S.~S. 1999,
  \apjs, 120, 51

\bibitem[{{Churchill} {et~al.}(2003){Churchill}, {Vogt}, \&
  {Charlton}}]{churchillvogtcharlton03}
{Churchill}, C.~W., {Vogt}, S.~S., \& {Charlton}, J.~C. 2003, \aj, 125, 98

\bibitem[{{de Heij} {et~al.}(2002){de Heij}, {Braun}, \&
  {Burton}}]{deheijbraunburton02}
{de Heij}, V., {Braun}, R., \& {Burton}, W.~B. 2002, \aap, 391, 159

\bibitem[{{Dekker} {et~al.}(2000){Dekker}, {D'Odorico}, {Kaufer}, {Delabre}, \&
  {Kotzlowski}}]{dekkeretal.00}
{Dekker}, H., {D'Odorico}, S., {Kaufer}, A., {Delabre}, B., \& {Kotzlowski}, H.
  2000, in Proc. SPIE Vol. 4008, p. 534-545, Optical and IR Telescope
  Instrumentation and Detectors, Masanori Iye; Alan F. Moorwood; Eds., ed.
  M.~{Iye} \& A.~F. {Moorwood}, 534--545

\bibitem[{{Ding} {et~al.}(2003{\natexlab{a}}){Ding}, {Charlton}, {Bond},
  {Zonak}, \& {Churchill}}]{dingetal03}
{Ding}, J., {Charlton}, J.~C., {Bond}, N.~A., {Zonak}, S.~G., \& {Churchill},
  C.~W. 2003{\natexlab{a}}, \apj, 587, 551

\bibitem[{{Ding} {et~al.}(2005){Ding}, {Charlton}, \&
  {Churchill}}]{dingcharltonchurchill05}
{Ding}, J., {Charlton}, J.~C., \& {Churchill}, C.~W. 2005, \apj, 621, 615

\bibitem[{{Ding} {et~al.}(2003{\natexlab{b}}){Ding}, {Charlton}, {Churchill},
  \& {Palma}}]{dingcharltonchurchillpalma03}
{Ding}, J., {Charlton}, J.~C., {Churchill}, C.~W., \& {Palma}, C.
  2003{\natexlab{b}}, \apj, 590, 746

\bibitem[{{Fraternali} {et~al.}(2007){Fraternali}, {Binney}, {Oosterloo}, \&
  {Sancisi}}]{fraternalietal_07}
{Fraternali}, F., {Binney}, J., {Oosterloo}, T., \& {Sancisi}, R. 2007, New
  Astronomy Review, 51, 95

\bibitem[{{Fraternali} \& {Binney}(2006)}]{fraternaliandbinney06}
{Fraternali}, F. \& {Binney}, J.~J. 2006, \mnras, 366, 449

\bibitem[{{Hartmann} \& {Burton}(1997)}]{hartmannburton97}
{Hartmann}, D. \& {Burton}, W.~B. 1997, {Atlas of Galactic Neutral Hydrogen}
  (Atlas of Galactic Neutral Hydrogen, by Dap Hartmann and W.~Butler Burton,
  pp.~243.~ISBN 0521471117.~Cambridge, UK: Cambridge University Press, February
  1997.)

\bibitem[{{Hoffman} {et~al.}(2004){Hoffman}, {Salpeter}, \&
  {Hirani}}]{hoffmanetal04}
{Hoffman}, G.~L., {Salpeter}, E.~E., \& {Hirani}, A. 2004, \aj, 128, 2932

\bibitem[{{Kalberla}(2003)}]{kalberla03}
{Kalberla}, P.~M.~W. 2003, \apj, 588, 805

\bibitem[{{Kalberla} {et~al.}(2005){Kalberla}, {Burton}, {Hartmann}, {Arnal},
  {Bajaja}, {Morras}, \& {P{\"o}ppel}}]{kalberlaetal05}
{Kalberla}, P.~M.~W., {Burton}, W.~B., {Hartmann}, D., {et~al.} 2005, \aap,
  440, 775

\bibitem[{{Kim} {et~al.}(2002){Kim}, {Carswell}, {Cristiani}, {D'Odorico}, \&
  {Giallongo}}]{kimcarswelletal02}
{Kim}, T.-S., {Carswell}, R.~F., {Cristiani}, S., {D'Odorico}, S., \&
  {Giallongo}, E. 2002, \mnras, 335, 555

\bibitem[{{Majewski}(2004)}]{majewski04}
{Majewski}, S. 2004, in Astronomical Society of the Pacific Conference Series,
  Vol. 327, Satellites and Tidal Streams, ed. F.~{Prada}, D.~{Martinez
  Delgado}, \& T.~J. {Mahoney}, 63--+

\bibitem[{{Masiero} {et~al.}(2005){Masiero}, {Charlton}, {Ding}, {Churchill},
  \& {Kacprzak}}]{masieroetal05}
{Masiero}, J.~R., {Charlton}, J.~C., {Ding}, J., {Churchill}, C.~W., \&
  {Kacprzak}, G. 2005, \apj, 623, 57

\bibitem[{{Mathewson} {et~al.}(1974){Mathewson}, {Cleary}, \&
  {Murray}}]{mathewson74}
{Mathewson}, D.~S., {Cleary}, M.~N., \& {Murray}, J.~D. 1974, \apj, 190, 291

\bibitem[{{Muller} {et~al.}(1963){Muller}, {Oort}, \&
  {Raimond}}]{mulleroortraimond63}
{Muller}, C.~A., {Oort}, J.~H., \& {Raimond}, E. 1963, C. R. Acad. Sc. Paris,
  257, 1661

\bibitem[{{M\"{u}nch}(1952)}]{muench52}
{M\"{u}nch}, G. 1952, \pasp, 64, 312

\bibitem[{{Petitjean} \& {Bergeron}(1990)}]{petitjeanbergeron90}
{Petitjean}, P. \& {Bergeron}, J. 1990, \aap, 231, 309

\bibitem[{{Petitjean} {et~al.}(1993){Petitjean}, {Webb}, {Rauch}, {Carswell},
  \& {Lanzetta}}]{petitjean93}
{Petitjean}, P., {Webb}, J.~K., {Rauch}, M., {Carswell}, R.~F., \& {Lanzetta},
  K. 1993, \mnras, 262, 499

\bibitem[{{Prochter} {et~al.}(2006){Prochter}, {Prochaska}, \&
  {Burles}}]{prochteretal06}
{Prochter}, G.~E., {Prochaska}, J.~X., \& {Burles}, S.~M. 2006, \apj, 639, 766

\bibitem[{{Putman} {et~al.}(2002){Putman}, {de Heij}, {Staveley-Smith},
  {Braun}, {Freeman}, {Gibson}, {Burton}, {Barnes}, {Banks}, {Bhathal}, {de
  Blok}, {Boyce}, {Disney}, {Drinkwater}, {Ekers}, {Henning}, {Jerjen},
  {Kilborn}, {Knezek}, {Koribalski}, {Malin}, {Marquarding}, {Minchin},
  {Mould}, {Oosterloo}, {Price}, {Ryder}, {Sadler}, {Stewart}, {Stootman},
  {Webster}, \& {Wright}}]{Putmanetal02}
{Putman}, M.~E., {de Heij}, V., {Staveley-Smith}, L., {et~al.} 2002, \aj, 123,
  873

\bibitem[{{Rao} {et~al.}(2006){Rao}, {Turnshek}, \& {Nestor}}]{raoetal06}
{Rao}, S.~M., {Turnshek}, D.~A., \& {Nestor}, D.~B. 2006, \apj, 636, 610

\bibitem[{{Richter}(2006)}]{richterp06}
{Richter}, P. 2006, in Reviews in Modern Astronomy, Vol.~19, Reviews in Modern
  Astronomy, ed. S.~{Roeser}, 31--+

\bibitem[{{Richter} {et~al.}(1999){Richter}, {de Boer}, {Widmann},
  {Kappelmann}, {Gringel}, {Grewing}, \& {Barnstedt}}]{richterdeboeretal99}
{Richter}, P., {de Boer}, K.~S., {Widmann}, H., {et~al.} 1999, \nat, 402, 386

\bibitem[{{Richter} {et~al.}(2003{\natexlab{a}}){Richter}, {Savage}, {Sembach},
  {Tripp}, \& {Jenkins}}]{richtersavagesembachetal03}
{Richter}, P., {Savage}, B.~D., {Sembach}, K.~R., {Tripp}, T.~M., \& {Jenkins},
  E.~B. 2003{\natexlab{a}}, in Astrophysics and Space Science Library, Vol.
  281, The IGM/Galaxy Connection. The Distribution of Baryons at z=0, ed. J.~L.
  {Rosenberg} \& M.~E. {Putman}, 85--+

\bibitem[{{Richter} {et~al.}(2003{\natexlab{b}}){Richter}, {Wakker}, {Savage},
  \& {Sembach}}]{richterwakkersavagesembach03}
{Richter}, P., {Wakker}, B.~P., {Savage}, B.~D., \& {Sembach}, K.~R.
  2003{\natexlab{b}}, \apj, 586, 230

\bibitem[{{Richter} {et~al.}(2005){Richter}, {Westmeier}, \&
  {Br{\"u}ns}}]{richterwestmeierbruens05}
{Richter}, P., {Westmeier}, T., \& {Br{\"u}ns}, C. 2005, \aap, 442, L49

\bibitem[{{Savage} \& {Massa}(1987)}]{savage_massa87}
{Savage}, B.~D. \& {Massa}, D. 1987, \apj, 314, 380

\bibitem[{{Savage} {et~al.}(2000){Savage}, {Wakker}, {Jannuzi}, {Bahcall},
  {Bergeron}, {Boksenberg}, {Hartig}, {Kirhakos}, {Murphy}, {Sargent},
  {Schneider}, {Turnshek}, \& {Wolfe}}]{savageetal00}
{Savage}, B.~D., {Wakker}, B., {Jannuzi}, B.~T., {et~al.} 2000, \apjs, 129, 563

\bibitem[{{Sembach} \& {Danks}(1994)}]{sembachdanks94S}
{Sembach}, K.~R. \& {Danks}, A.~C. 1994, \aap, 289, 539

\bibitem[{{Sembach} {et~al.}(1993){Sembach}, {Danks}, \&
  {Savage}}]{sembachdankssavage93}
{Sembach}, K.~R., {Danks}, A.~C., \& {Savage}, B.~D. 1993, \aaps, 100, 107

\bibitem[{{Sembach} \& {Savage}(1996)}]{sembachsavage96}
{Sembach}, K.~R. \& {Savage}, B.~D. 1996, \apj, 457, 211

\bibitem[{{Sembach} {et~al.}(1991){Sembach}, {Savage}, \&
  {Massa}}]{Sembachetal91}
{Sembach}, K.~R., {Savage}, B.~D., \& {Massa}, D. 1991, \apj, 372, 81

\bibitem[{{Sembach} {et~al.}(2003{\natexlab{a}}){Sembach}, {Wakker}, {Savage},
  {Richter}, {Meade}, {Shull}, {Jenkins}, {Sonneborn}, \&
  {Moos}}]{sembach_wakker_savage_richter_etal03}
{Sembach}, K.~R., {Wakker}, B.~P., {Savage}, B.~D., {et~al.}
  2003{\natexlab{a}}, \apjs, 146, 165

\bibitem[{{Sembach} {et~al.}(2003{\natexlab{b}}){Sembach}, {Wakker}, {Savage},
  {Richter}, {Meade}, {Shull}, {Jenkins}, {Sonneborn}, \& {Moos}}]{sembach03}
{Sembach}, K.~R., {Wakker}, B.~P., {Savage}, B.~D., {et~al.}
  2003{\natexlab{b}}, \apjs, 146, 165

\bibitem[{{Shapiro} \& {Benjamin}(1991)}]{shapiro_benjamin91}
{Shapiro}, P.~R. \& {Benjamin}, R.~A. 1991, \pasp, 103, 923

\bibitem[{{Shapiro} \& {Field}(1976)}]{shapirofield76}
{Shapiro}, P.~R. \& {Field}, G.~B. 1976, \apj, 205, 762

\bibitem[{{Spitzer}(1956)}]{spitzer56}
{Spitzer}, L.~J. 1956, \apj, 124, 20

\bibitem[{{Thom} {et~al.}(2006){Thom}, {Putman}, {Gibson}, {Christlieb},
  {Flynn}, {Beers}, {Wilhelm}, \& {Lee}}]{thom2006}
{Thom}, C., {Putman}, M.~E., {Gibson}, B.~K., {et~al.} 2006, \apjl, 638, L97

\bibitem[{{van Woerden} {et~al.}(1999){van Woerden}, {Schwarz}, {Peletier},
  {Wakker}, \& {Kalberla}}]{vanWoerden99}
{van Woerden}, H., {Schwarz}, U.~J., {Peletier}, R.~F., {Wakker}, B.~P., \&
  {Kalberla}, P.~M.~W. 1999, \nat, 400, 138

\bibitem[{{Wakker}(1991)}]{wakker91}
{Wakker}, B.~P. 1991, \aap, 250, 499

\bibitem[{{Wakker}(2001)}]{wakker01}
{Wakker}, B.~P. 2001, \apjs, 136, 463

\bibitem[{{Wakker} \& {Mathis}(2000)}]{wakkermathis00}
{Wakker}, B.~P. \& {Mathis}, J.~S. 2000, \apjl, 544, L107

\bibitem[{{Wakker} {et~al.}(2007){Wakker}, {York}, {Howk}, {Barentine},
  {Wilhelm}, {Peletier}, {van Woerden}, {Beers}, {Ivezi{\'c}}, {Richter}, \&
  {Schwarz}}]{wakker_york_howketal07}
{Wakker}, B.~P., {York}, D.~G., {Howk}, J.~C., {et~al.} 2007, \apjl, 670, L113

\bibitem[{{Wakker} {et~al.}(2008){Wakker}, {York}, {Wilhelm}, {Barentine},
  {Richter}, {Beers}, {Ivezi{\'c}}, \& {Howk}}]{wakkeryorkwilhelmetal08}
{Wakker}, B.~P., {York}, D.~G., {Wilhelm}, R., {et~al.} 2008, \apj, 672, 298

\bibitem[{{Westmeier}(2007)}]{westmeier07}
{Westmeier}, T. 2007, PhD thesis, {Rheinische
  Friedrich-Wilhelms-Universit\"{a}t Bonn}

\end{thebibliography}

\begin{figure*} 9067ref.bib
\centering 
\includegraphics[width=0.9\textwidth,bb=90 128 505 715,clip=]{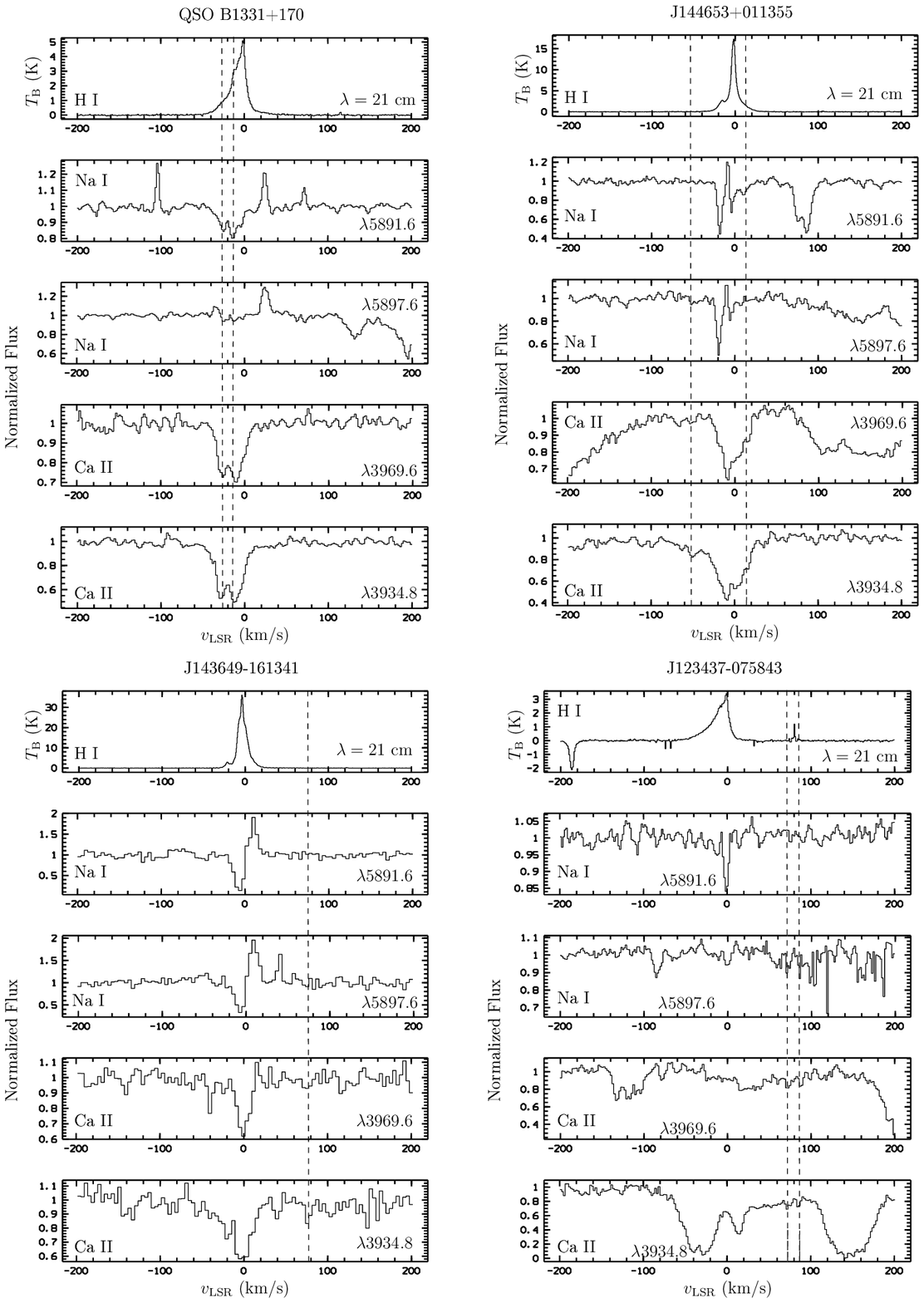}
\caption{\ion{Ca}{ii} and \ion{Na}{i} absorption spectra of QSO\,B1331+170$+$2411 J144653$+$011355, J143649$-$161341, and 
J123437$-$075843 obtained with UVES. Additionally the corresponding \ion{H}{i} emission line profiles measured with the Effelsberg 100-m telescope are shown. The spikes in the Effelsberg spectra of J123437$-$075843 are radio frequency interferences (RFI).}
\label{fig_QSOB133_J144653_J143649_J123437} 
\end{figure*} 

\begin{figure*} 
\centering 
\includegraphics[width=0.9\textwidth,bb=90 128 505 715,clip=]{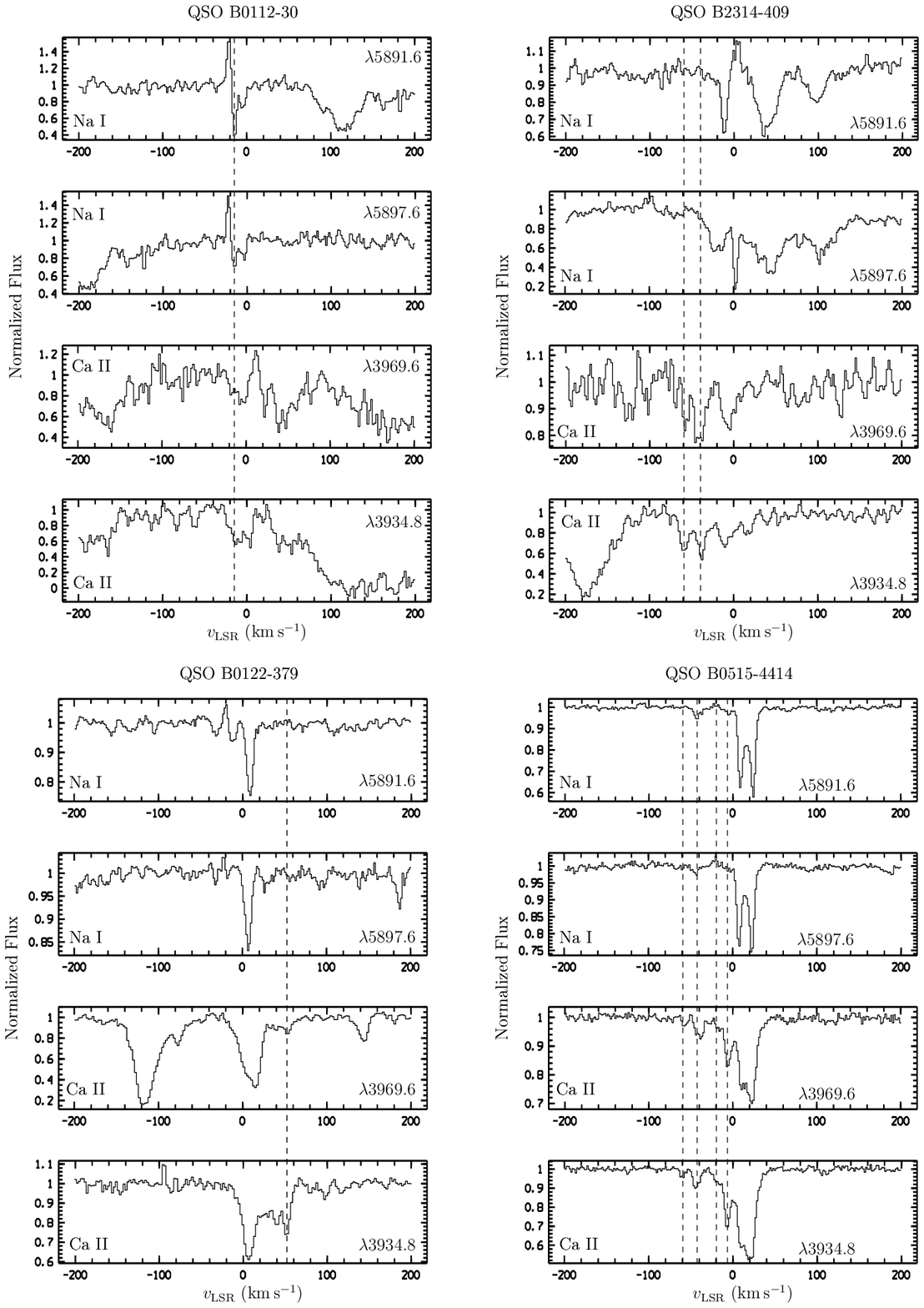}
\caption{\ion{Ca}{ii} and \ion{Na}{i} absorption spectra of QSO\,B0112$-$30, QSO\,B2314$-$409, QSO\,B0122$-$379, and QSO\,B0515$-$4414 obtained with UVES. For these sight lines no Effelsberg data are available.} 
\label{fig_QSOB0112-30_QSOB2314_q0122_QSOB0515} 
\end{figure*}

\begin{figure*} 
\centering 
\includegraphics[width=0.9\textwidth,bb=90 128 505 715,clip=]{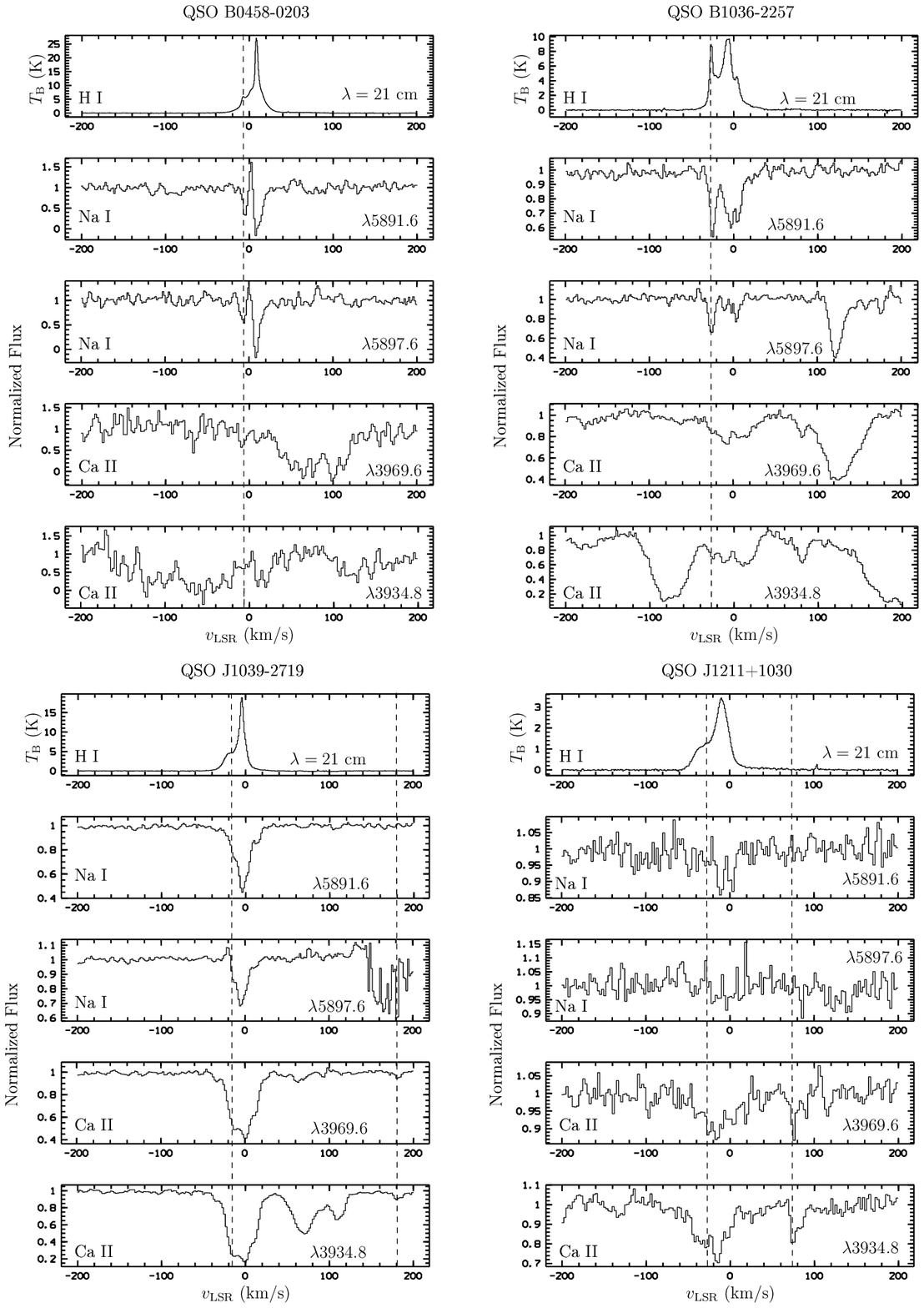}
\caption{\ion{Ca}{ii} and \ion{Na}{i} absorption spectra of QSO\,B0458$-$0203, QSO\,B1036$-$2257, QSO\,J1039$-$2719, and QSO\,J1211$+$1030 obtained with UVES. Additionally the corresponding \ion{H}{i} emission line profiles measured with the Effelsberg 100-m telescope are shown.} 
\label{fig_QSOB0458_QSOB1036} 
\end{figure*}

\begin{figure*} 
\centering 
\includegraphics[width=0.9\textwidth,bb=90 128 505 715,clip=]{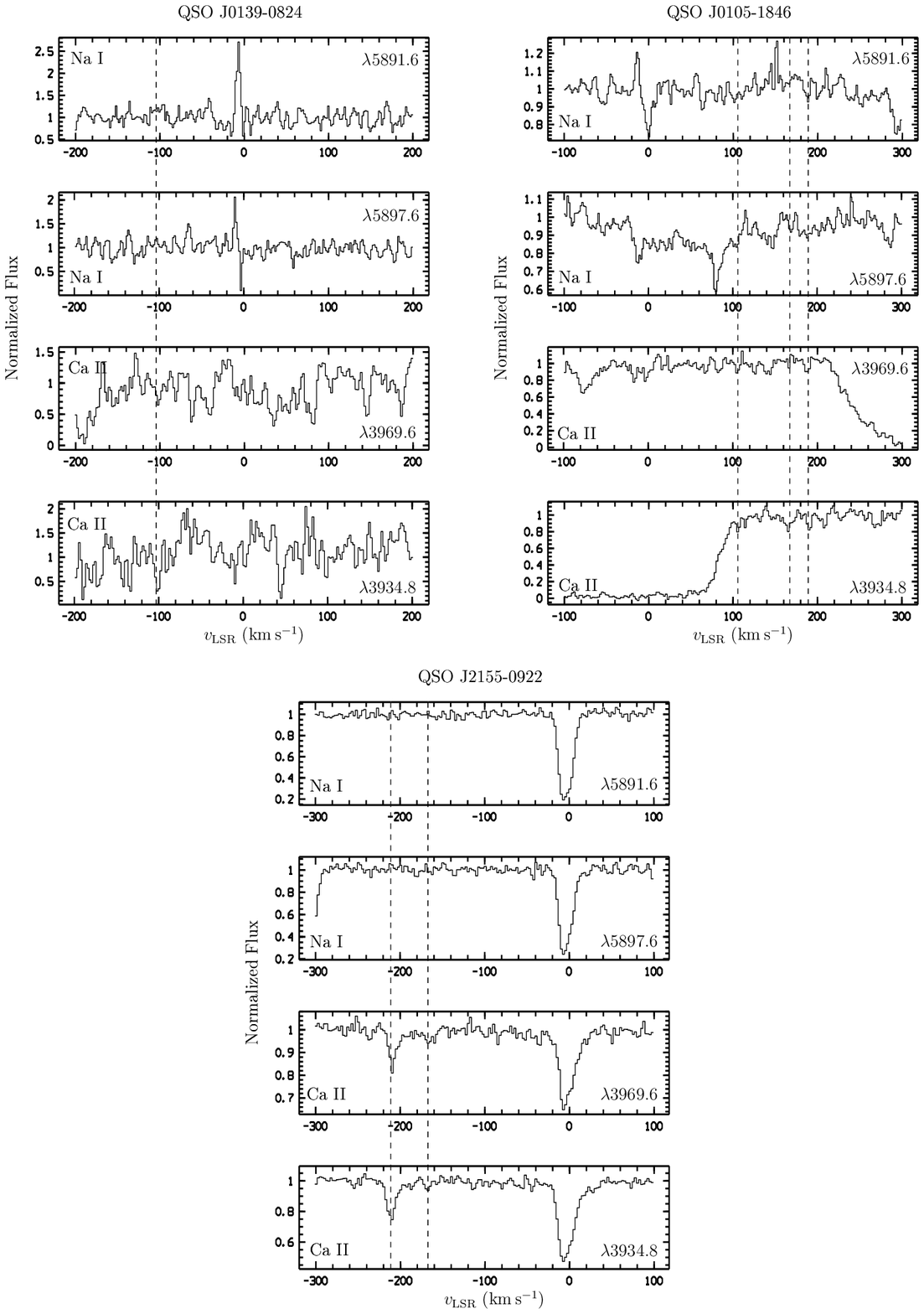}
\caption{\ion{Ca}{ii} and \ion{Na}{i} absorption spectra of QSO\,J0139$-$0824, QSO\,J0105$-$1846, and QSO\,J2155$-$0922 obtained with UVES. No Effelsberg data are available.} 
\label{fig_QSOJ0139_QSOJ0105_QSOJ2155_0922_} 
\end{figure*}

\begin{figure*} 
\centering 
\includegraphics[width=0.9\textwidth,bb=90 233 505 608,clip=]{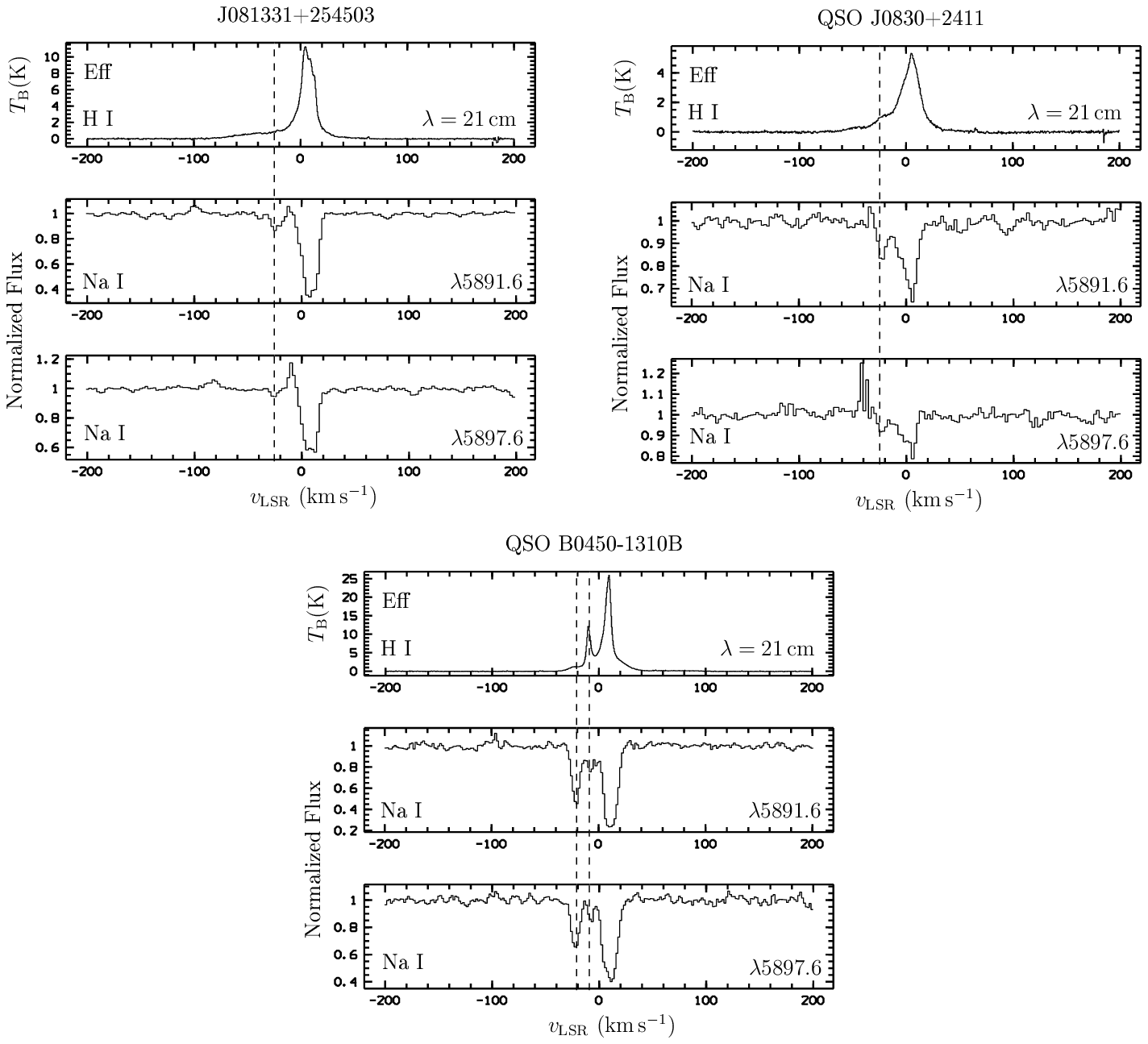}
\caption{\ion{Na}{i} absorption and \ion{H}{i} emission spectra in the direction of the quasars J081331$+$254503, QSO\,J0830$+$2411, and QSO\,B0450$-$1310B obtained with UVES and the Effelsberg 100-m telescope. For these sight lines no \ion{Ca}{ii} data are available.} 
\label{fig_J081331_QSOJ0830_QSOB0450} 
\end{figure*}

\begin{figure*} 
\centering 
\includegraphics[width=0.9\textwidth,bb=90 218 505 623,clip=]{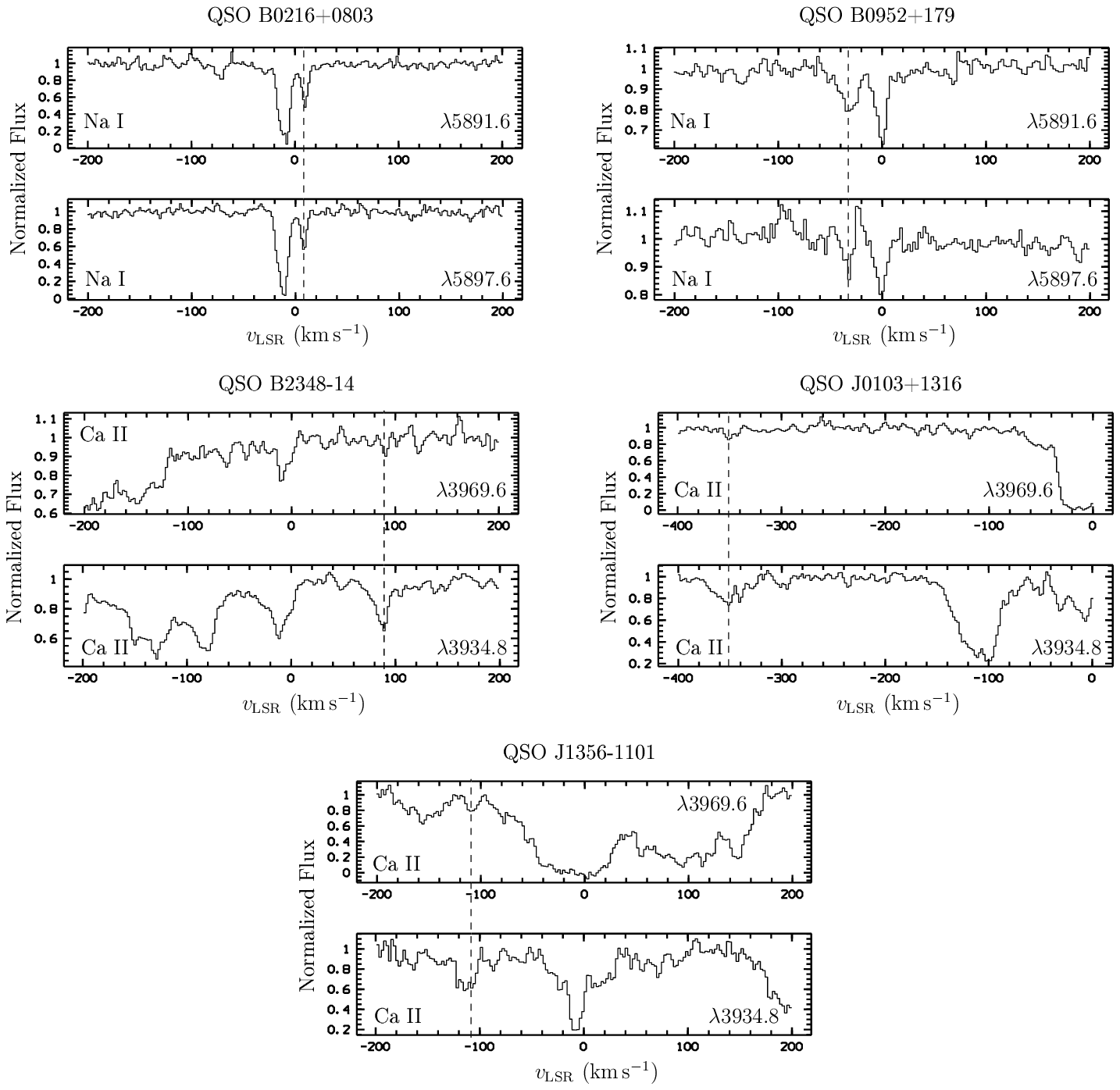}
\caption{\textbf{Upper row}: \ion{Na}{i} absorption spectra of QSO\,B0216$+$0803 and QSO\,B0952$+$179 obtained with UVES. There are no \ion{Ca}{ii} data available. \textbf{Middle and lower rows}: \ion{Ca}{ii} absorption spectra of QSO\,B2348$-$147,QSO\,J0103$+$1316, and QSO\,J1356$-$1101  obtained with UVES. There are no \ion{Na}{i} data available. For all five sight lines we have no Effelsberg data.} 
\label{fig_QSOB0216_QSOB0952_QSOB2348_QSOJ0103_QSOJ1356} 
\end{figure*}

\end{document}